\documentclass[twoside, onecolumn, 11pt]{IEEEtran}

\usepackage{epsfig}
\usepackage{graphicx}
\usepackage{amsmath}
\usepackage{amssymb}
\usepackage{float}
\usepackage{url}
\newtheorem{definition}{Definition}
\newtheorem{lemma}{Lemma}
\newtheorem{proposition}{Proposition}
\newtheorem{corollary}{Corollary}
\newtheorem{theorem}{Theorem}

\newtheorem{remark}{Remark}
\newtheorem{example}{Example}

\newcommand{\dsum}{\displaystyle\sum}

\newcommand{\naturals}{\ensuremath{\mathbb{N}}}
\newcommand{\reals}{\ensuremath{\mathbb{R}}}
\newcommand{\pr}{\ensuremath{\mathbb{P}}}
\newcommand{\expectation}{\ensuremath{\mathbb{E}}}

\begin{document}

\title{Entropy Bounds for Discrete Random Variables via Maximal Coupling}

\markboth{Final Version. Date: July 23, 2013. Accepted for publication in the
IEEE Transactions on Information Theory.}
{I. SASON: Entropy Bounds for Discrete Random Variables via Maximal Coupling}

\author{Igal Sason,~\IEEEmembership{Senior Member, IEEE}
\thanks{\copyright 2013 IEEE. Personal use of this material is permitted.
However, permission to use this material for any other purposes must be obtained
from the IEEE by sending a request to pubs-permissions@ieee.org.}
\thanks{The manuscript was submitted to the {\em IEEE Trans. on Information
Theory} in September 20, 2012, and accepted in July 18, 2013.
This research work was supported by the
Israel Science Foundation (grant no. 12/12). The material in
this paper was presented in part at the 2013 International
Symposium on Information Theory (ISIT '13), Istanbul, Turkey,
July 2013.}
\thanks{Communicated by Ioannis Kontoyiannis, Associate Editor At Large.}
\thanks{The author is with the Department of Electrical Engineering,
Technion--Israel Institute of Technology, Haifa 32000, Israel.
His e-mail address is sason@ee.technion.ac.il.}}

\maketitle
\date{today}

\begin{abstract}
This paper derives new bounds on the difference of the
entropies of two discrete random variables in terms of the local and
total variation distances between their probability mass functions.
The derivation of the bounds relies on maximal coupling,
and they apply to discrete random variables which are defined
over finite or countably infinite alphabets. Loosened versions of these
bounds are demonstrated to reproduce some previously reported results.
The use of the new bounds is exemplified for the Poisson approximation,
where bounds on the local and total variation distances follow
from Stein's method.
\end{abstract}


\vspace*{-0.2cm}
\section{Introduction}
\label{section: Introduction}
The question of quantifying the continuity (or lack of it) of entropy,
with respect to natural topologies on discrete probability distributions
is fundamental. This question has been studied in the literature for the
topology induced by the total variation distance, and there it is well known
that the entropy is continuous when the alphabet is finite, but it is
not necessarily continuous when the alphabet is countably infinite.
The interplay between the difference of the entropies of two discrete random
variables and their total variation distance has been extensively studied
(see, e.g., \cite[Theorem~17.3.3]{Cover_Thomas}, \cite{Csiszar_Korner_book},
\cite[Lemma~1]{Csiszar_PPI1996},
\cite{Harremoes_2003}--\cite{entropy_difference_and_variational_distance_IT2010},
\cite{Kovacevic_arxiv2013}, \cite{Pinsker_PPI2005}--\cite{Prelov_PPI2010},
\cite{Sason_ITW}, \cite{Topsoe_IT2000}, \cite{Zhang_IT2007}).

New bounds on the difference of the entropies of two discrete random variables
are derived in this work. The bounds apply to random variables with finite or countably
infinite alphabet, and they improve some previously reported bounds. The derivation of
the new bounds relies on the notion of {\em maximal coupling}, which is also known to
be useful for the derivation of error bounds via Stein's method (see, e.g.,
\cite[Chapter~2]{RossP_book07} and \cite{Ross_Tutorial11}). Stein's method also serves
to exemplify the use of the new bounds in the context of the Poisson approximation.
The link between Stein's method and information theory was pioneered in \cite{BarbourJKM_EJP_2010}
in the context of the compound Poisson approximation, and a
recent work \cite{Ley_Swan_arxiv2012} (that was done independently and in parallel to this work)
further links between information theory and Stein's method for discrete probability distributions.

To set definitions and notation, we introduce essential terms that
serve to derive the new bounds in this paper.

\begin{definition}
A {\em coupling} of a pair of two random variables $(X, Y)$
is a pair of two random variables $(\hat{X}, \hat{Y})$ with the same
marginal probability distributions as of $(X, Y)$.
\label{definition: coupling}
\end{definition}

\begin{definition}
For a pair of random variables $(X, Y)$, a coupling $(\hat{X}, \hat{Y})$
is called a {\em maximal coupling} if $\pr(\hat{X}=\hat{Y})$ gets its maximal
value among all the couplings of $(X,Y)$.
\label{definition: maximal coupling}
\end{definition}

\begin{definition}
Let $X$ and $Y$ be discrete random variables that take values
in a set $\mathcal{A}$, and let $P_X$ and $P_Y$ be their respective probability
mass functions. The {\em local distance} and {\em total variation distance}
between $X$ and $Y$ are, respectively,
\begin{eqnarray}
&& d_{\text{loc}}(X,Y) \triangleq \sup_{u \in \mathcal{A}} |P_X(u) - P_Y(u)|
\label{eq: local distance} \\[0.1cm]
&& d_{\text{TV}}(X,Y) \triangleq \frac{1}{2} \, \sum_{u \in \mathcal{A}} |P_X(u) - P_Y(u)|.
\label{eq: total variation distance}
\end{eqnarray}
The local distance is the $l^{\infty}$ distance
between the probability mass functions, and the total variation
distance is half the $l^1$ distance.
The factor of one-half on the right-hand side of \eqref{eq: total variation distance}
normalizes the total variation distance to get values between zero and one. It is noted
that the notation in the literature is not consistent, with a factor~2 on the right-hand
side of \eqref{eq: total variation distance} often being present or not.
It is easy to show (see, e.g., \cite[Lemma~5.4 on pp.~133--134]{Gray}) that with this definition
$$d_{\text{TV}}(X,Y) = \sup_{\mathcal{B} \subseteq \mathcal{A}}
|\pr(X \in \mathcal{B}) - \pr(Y \in \mathcal{B})|.$$
From the last equality and the definition of the local distance in
\eqref{eq: local distance}, it follows that
$d_{\text{loc}}(X,Y) \leq d_{\text{TV}}(X,Y).$

A basic property that links between maximal coupling
and the total variation distance is that if $(\hat{X}, \hat{Y})$ is a
maximal coupling of $(X,Y)$ then $\pr(\hat{X} \neq \hat{Y}) = d_{\text{TV}}(X,Y)$.
Throughout this paper, the term `distribution' refers to the probability mass function
of a discrete random variable defined over a finite or countably infinite alphabet.
\label{definition: local and total variation distances}
\end{definition}

The following theorem is a basic result on maximal coupling that also
suggests, as part of its proof, a construction for maximal coupling
(see, e.g., \cite[Chapter~2]{RossP_book07}).
We later rely on this particular construction to derive in
Section~\ref{section: Refined Bounds on the Entropy of Discrete Random Variables via Coupling}
some new bounds on the entropy of discrete random variables.

\begin{theorem}
Let $X$ and $Y$ be discrete random variables that take values
in a set $\mathcal{A}$, and let their respective probability mass functions be
$$P_X(x) = \pr(X=x), \quad P_Y(y) = \pr(Y=y), \quad \forall \, x, y \in \mathcal{A}.$$
Then, the maximal coupling of $(X,Y)$ satisfies
\begin{equation}
\pr(\hat{X} = \hat{Y}) = \sum_{u \in \mathcal{A}} \min \{ P_X(u), P_Y(u) \}.
\label{eq: maximal coupling}
\end{equation}
\label{theorem: maximal coupling}
\end{theorem}
\vspace*{-0.4cm}
\begin{proof}
Let $\mathcal{B} \triangleq \{u \in \mathcal{A}: \, P_X(u) < P_Y(u) \}$, and
let $\mathcal{B}^{\text{c}} \triangleq \mathcal{A} \setminus \mathcal{B}$.
Then, for every coupling $(\hat{X}, \hat{Y})$ of $(X, Y)$,
\begin{eqnarray}
&& \pr(\hat{X} = \hat{Y}) \nonumber \\
&& = \pr(\hat{X} = \hat{Y}, \, \hat{Y} \in \mathcal{B}) +
\pr(\hat{X} = \hat{Y}, \, \hat{Y} \in \mathcal{B}^{\text{c}})
\nonumber \\
&& \leq \pr(\hat{X} \in \mathcal{B}) + \pr(\hat{Y} \in \mathcal{B}^{\text{c}}) \nonumber \\
&& = \pr(X \in \mathcal{B}) + \pr(Y \in \mathcal{B}^{\text{c}}) \nonumber \\
&& = \sum_{u \in \mathcal{B}} P_X(u) + \sum_{u \in \mathcal{B}^{\text{c}}} P_Y(u) \nonumber \\
&& = \sum_{u \in \mathcal{B}} \min \{P_X(u), P_Y(u)\}
+ \sum_{u \in \mathcal{B}^{\text{c}}} \min\{P_X(u), P_Y(u)\} \nonumber \\
&& = \sum_{u \in \mathcal{A}} \min \{P_X(u), P_Y(u)\} \triangleq p.
\label{eq: upper bound on the probability that a coupled pair is equal to each other}
\end{eqnarray}

The following provides a construction of a coupling $(\hat{X}, \hat{Y})$ that achieves
the bound in \eqref{eq: upper bound on the probability that a coupled pair is equal to each other}
with equality, so it forms a maximal coupling of $(X,Y)$.
Let $U$, $V$, $W$ and $J$ be independent discrete random variables, where
\begin{equation}
\pr(J=0) = 1-p, \quad \pr(J=1) = p
\label{eq: probability mass function of J}
\end{equation}
so $J \sim \text{Bernoulli}(p)$, and let $U$, $V$, $W$ have the following
probability mass functions:
\begin{eqnarray}
&& \hspace*{-1.2cm}
P_U(u) = \frac{\min \{P_X(u), P_Y(u)\}}{p}, \quad \forall \, u \in \mathcal{A}
\label{eq: probability mass function of U} \\
&& \hspace*{-1.2cm}
P_V(v) = \frac{P_X(v) - \min\{P_X(v), P_Y(v)\}}{1-p}, \quad \forall \, v \in \mathcal{A}
\label{eq: probability mass function of V} \\
&& \hspace*{-1.2cm}
P_W(w) = \frac{P_Y(w) - \min\{P_X(w), P_Y(w)\}}{1-p}, \quad \forall \, w \in \mathcal{A}.
\label{eq: probability mass function of W}
\end{eqnarray}
If $J=1$, let $\hat{X} = \hat{Y} = U$, and if $J=0$ let $\hat{X}=V$ and $\hat{Y}=W$.
For every $x, y \in \mathcal{A}$
\begin{eqnarray*}
&& \hspace*{-1cm} P_{\hat{X}}(x) \\
&& \hspace*{-1cm} = p \, \pr(\hat{X} = x \, | \, J=1) + (1-p) \, \pr(\hat{X} = x \, | \, J=0) \\
&& \hspace*{-1cm} = p \, P_U(x) + (1-p) \, P_V(x) \\
&& \hspace*{-1cm} = P_X(x)
\end{eqnarray*}
and similarly $P_{\hat{Y}}(y) = P_Y(y)$, so $(\hat{X}, \hat{Y})$ is indeed
a coupling of $(X,Y)$. Furthermore,
\begin{equation}
\pr(\hat{X} = \hat{Y}) \geq \pr(J=1) = p
\label{eq: lower bound on the probability that a coupled pair is equal to each other}
\end{equation}
so, from \eqref{eq: upper bound on the probability that a coupled pair is equal to each other}
and \eqref{eq: lower bound on the probability that a coupled pair is equal to each other},
it follows that the proposed construction for $(\hat{X}, \hat{Y})$ forms a maximal
coupling of $(X, Y)$ and also $\pr(\hat{X} = \hat{Y}) = p$.
\end{proof}

The following result is a simple consequence of Theorem~\ref{theorem: maximal coupling}
(see, e.g., \cite[Chapter~2]{RossP_book07}),
and it is also used for the derivation of the new bounds on the entropy in
Section~\ref{section: Refined Bounds on the Entropy of Discrete Random Variables via Coupling}.
\begin{theorem}
Let $X$ and $Y$ be two discrete random variables that take values in a set
$\mathcal{A}$. If $(\hat{X}, \hat{Y})$ is a maximal coupling of $(X, Y)$ then
\begin{equation}
\pr(\hat{X} \neq \hat{Y}) = d_{\text{TV}}(X, Y).
\label{eq: maximal coupling - second equality}
\end{equation}
\label{theorem: maximal coupling - second result}
\end{theorem}
\vspace*{-0.4cm}
\begin{proof}
This follows from \eqref{eq: total variation distance} and \eqref{eq: maximal coupling},
and the equality
$\min\{a, b\} = \frac{a+b-|a-b|}{2}$ for all $a,b \in \reals.$
\end{proof}

This work refines bounds on the difference of the entropies of two discrete random variables
via the use of maximal couplings, leading to sharpened bounds
that depend on both the local and total variation distances. The reader is also referred to
a recent work in \cite{Kovacevic_arxiv2013} that derived bounds for information measures
by relying on the notion of the {\em minimum entropy coupling}.

The main observation of this work is that if the local distance between two probability
distributions on a finite alphabet is smaller than the total variation distance, then
the bounds on the entropy difference can be significantly strengthened.
The second observation made
in this work is that there is an extension of the new bound to countably infinite alphabets, where
just knowing the total variation distance between two distributions does not imply anything
about the difference of the respective entropies.
The new bound that follows from the second observation is applied in this work to obtain refined bounds
on the entropy of sums of independent (possibly non-identically distributed) Bernoulli random variables
that arise in numerous applications. The application of the new bounds
to the Poisson approximation is facilitated by using
bounds on the total variation and local distances which follow from Stein's method, and
the improvement that is obtained by these bounds is exemplified in this work.
For comparison, a looser version of the new bounds was earlier applied in
\cite{Sason_ITW} to get bounds on the entropy of sums of
dependent and non-identically distributed Bernoulli random variables.

The continuation of this paper is structured as follows:
Section~\ref{section: A Proof of a Known Bound on the Entropy of Discrete Random Variables via Coupling}
introduces a known bound, due to Zhang \cite{Zhang_IT2007}, on the difference of the entropies
of two discrete random variables in terms of the total variation distance.
A shortened proof that is based on maximal coupling serves to motivate the derivation
of some refined bounds. These new bounds, proved in
Section~\ref{section: Refined Bounds on the Entropy of Discrete Random Variables via Coupling}
via maximal coupling, depend on both the local and total variation distances.
Section~\ref{section: Examples} exemplifies the use of the new bounds with a link to
Stein's method, and it also compares them with some previously known bounds. Finally, the paper is
concluded in Section~\ref{section: summary}. Throughout this paper, the logarithms and the entropies
are to the base~$e$.

\section{A Proof of a Known Bound on the Entropy of Discrete Random Variables via Coupling}
\label{section: A Proof of a Known Bound on the Entropy of Discrete Random Variables via Coupling}
The following theorem relies on a bound that first appeared in
\cite[Eq.~(4)]{Zhang_IT2007} and proved by coupling. It was
later introduced in \cite[Theorem~6]{entropy_difference_and_variational_distance_IT2010}
by re-proving the inequality in a different way (without coupling), and it
was also strengthened there by showing an explicit case where the following bound is tight.
As is proved in \cite[Section~3]{Zhang_IT2007}, the bound on the entropy difference that
is introduced in the following theorem improves the bound in \cite[Theorem~17.3.3]{Cover_Thomas}
or \cite[Lemma~2.7]{Csiszar_Korner_book}.

\begin{theorem}
Let $X$ and $Y$ be two discrete random variables that take values in a finite set $\mathcal{A}$,
and let $|\mathcal{A}| = M$. Then,
\begin{eqnarray}
|H(X) - H(Y)| \leq d_{\text{TV}}(X,Y) \, \log(M-1) + h\bigl(d_{\text{TV}}(X,Y)\bigr)
\label{eq: Zhang's bound}
\end{eqnarray}
where $h$ denotes the binary entropy function. Furthermore, there is a case where the
bound is tight.
\label{theorem: known bound on the entropy difference in terms of total variation distance}
\end{theorem}

The following proof of
Theorem~\ref{theorem: known bound on the entropy difference in terms of total variation distance}
exemplifies the use of maximal coupling in proving an information-theoretic result.

\begin{proof}
Let $(\hat{X}, \hat{Y})$ be a maximal coupling of $(X,Y)$.
Since $H(X) = H(\hat{X})$ and $H(Y) = H(\hat{Y})$ (note that the marginal probability
mass functions of $(X,Y)$ and $(\hat{X}, \hat{Y})$ are the same), it follows from
Fano's inequality and Theorem~\ref{theorem: maximal coupling - second result}
(see~\eqref{eq: maximal coupling - second equality}) that
\begin{eqnarray*}
&& \hspace*{-0.2cm} \bigl| H(X) - H(Y) \bigr| \\
&& = \bigl| H(\hat{X}) - H(\hat{Y}) \bigr| \\
&& = \bigl| H(\hat{X} | \hat{Y}) - H(\hat{Y} | \hat{X}) \bigr| \\
&& \leq \max \Bigl\{ H(\hat{X} | \hat{Y}), H(\hat{Y} | \hat{X}) \Bigr\} \\
&& \leq \pr(\hat{X} \neq \hat{Y}) \, \log(M-1) + h\bigl( \pr(\hat{X} \neq \hat{Y}) \bigr) \\
&& = d_{\text{TV}}(X,Y) \, \log(M-1) + h\bigl( d_{\text{TV}}(X,Y) \bigr).
\end{eqnarray*}
This proves the bound in \eqref{eq: Zhang's bound} (see \cite[Eq.~(4)]{Zhang_IT2007}).
If $d_{\text{TV}}(X,Y) \leq \varepsilon$ for some
$\varepsilon \in \bigl[0, 1 - \frac{1}{M}\bigr]$,
the replacement of $d_{\text{TV}}(X,Y)$ in the last bound
by $\varepsilon$ is valid; this holds since the function
$f(x) \triangleq x \log(M-1) + h(x)$ is monotonic increasing
over the interval $[0, 1-\frac{1}{M}]$ (since
$f'(x) = \log(M-1) + \log\left(\frac{1-x}{x}\right) > 0$ for
$0<x<1-\frac{1}{M}$). Otherwise, if $\varepsilon > 1 - \frac{1}{M}$,
$$\bigl| H(X) - H(Y) \bigr| \leq \max \Bigl\{H(X), H(Y) \Bigr\} \leq \log(M).$$

{\em Cases where the bound is tight \cite{entropy_difference_and_variational_distance_IT2010}}:
If $\varepsilon \in [0, 1-\frac{1}{M}]$, the bound is tight when
\begin{eqnarray*}
&& X \sim P_X = \left(1-\varepsilon, \frac{\varepsilon}{M-1},
\ldots, \frac{\varepsilon}{M-1} \right) \\
&& Y \sim P_Y = (1, 0, \ldots, 0)
\end{eqnarray*}
which implies that
\begin{eqnarray*}
&& d_{\text{TV}}(X,Y) = \varepsilon, \\
&& |H(X) - H(Y)| = H(X) = h(\varepsilon) + \varepsilon \log(M-1).
\end{eqnarray*}

If $\varepsilon \in (1-\frac{1}{M}, 1]$ then
the bound is tight when
\begin{eqnarray*}
X \sim \Bigl(\frac{1}{M}, \ldots, \frac{1}{M}\Bigr), \quad
Y \sim (1, 0, \ldots, 0)
\end{eqnarray*}
so, $d_{\text{TV}}(X,Y) = 1 - \frac{1}{M} < \varepsilon$
and $|H(X)-H(Y)| = \log(M)$.
\end{proof}

\section{New Bounds on the Entropy of Discrete Random Variables via Coupling}
\label{section: Refined Bounds on the Entropy of Discrete Random Variables via Coupling}

In the cases where the known bound in
Theorem~\ref{theorem: known bound on the entropy difference in terms of total variation distance}
was shown to be tight in \cite{entropy_difference_and_variational_distance_IT2010} (see the last
part of the proof in Section~\ref{section: A Proof of a Known Bound on the Entropy of Discrete Random Variables via Coupling}),
it is easy to verify that the local distance is equal to the total variation distance. However,
as is shown in the following, if it is not the case (i.e., the local
distance is smaller than the total variation distance), then the bound in
Theorem~\ref{theorem: known bound on the entropy difference in terms of total variation distance}
is {\em necessarily} not tight. Furthermore, this section provides new bounds
that depend on both the total variation and local distances.
If these two distances are equal then the new bound is particularized to the bound in
Theorem~\ref{theorem: known bound on the entropy difference in terms of total variation distance}
but otherwise, the new bound improves the bound in
Theorem~\ref{theorem: known bound on the entropy difference in terms of total variation distance}.
The general approach for proving the following new inequalities relies on the
construction of the maximal coupling that is introduced in the proof of
Theorem~\ref{theorem: maximal coupling}. The new results are stated and
proved in the following.

\begin{theorem}
Let $X$ and $Y$ be two discrete random variables that take values in a
finite set $\mathcal{A}$, and let $|\mathcal{A}| = M$. Then,
\begin{eqnarray}
|H(X)-H(Y)| \leq d_{\text{TV}}(X,Y) \, \log(M\alpha-1) + h\bigl(d_{\text{TV}}(X,Y)\bigr)
\label{eq: 1st new bound}
\end{eqnarray}
where
\begin{equation}
\alpha \triangleq \frac{d_{\text{loc}}(X,Y)}{d_{\text{TV}}(X,Y)}
\label{eq: alpha}
\end{equation}
denotes the ratio of the local and total variation distances (so,
$\alpha \in [\frac{2}{M}, 1]$), and $h$ denotes the binary entropy function.
Furthermore, if the probability mass functions of $X$ and $Y$ satisfy the
condition that $\frac{1}{2} \leq \frac{P_X}{P_Y} \leq 2$ whenever $P_X, P_Y > 0$,
then the bound in \eqref{eq: 1st new bound} is tightened to
\begin{eqnarray}
|H(X)-H(Y)| \leq d_{\text{TV}}(X,Y) \, \log\left(\frac{M\alpha-1}{4}\right) +
h\bigl(d_{\text{TV}}(X,Y)\bigr).
\label{eq: possible refinement of 1st new bound}
\end{eqnarray}
\label{theorem: 1st new bound}
\end{theorem}

\begin{remark}
Since, in general, $\alpha \leq 1$ then the case where $\alpha = 1$ is the worst case
for the bound in \eqref{eq: 1st new bound}. In the latter case, it is particularized to the
bound in
Theorem~\ref{theorem: known bound on the entropy difference in terms of total variation distance}
(see \cite[Theorem~6]{entropy_difference_and_variational_distance_IT2010} or
\cite[Eq.~(4)]{Zhang_IT2007}).
\end{remark}

\hspace*{0.1cm}
\begin{remark}
If $\alpha \leq \frac{1}{N}$ for some integer $N$
(since $\alpha \in \bigl[\frac{2}{M}, 1 \bigr]$ then
$N \in \{1, \ldots, \lfloor \frac{M}{2} \rfloor \}$),
the bound in \eqref{eq: 1st new bound} implies that
\begin{eqnarray}
|H(X) - H(Y)| \leq d_{\text{TV}}(X,Y) \, \log\left(\frac{M-N}{N}\right) + h\bigl(d_{\text{TV}}(X,Y)\bigr).
\label{eq: special case of 1st new bound}
\end{eqnarray}
The bounds in \eqref{eq: special case of 1st new bound} and \cite[Theorem~7]{entropy_difference_and_variational_distance_IT2010} are
similar {\em but they hold under different conditions}. The
bound in \cite[Theorem~7]{entropy_difference_and_variational_distance_IT2010}
requires that $P_X, P_Y \leq \frac{1}{N}$ everywhere, whereas the bound
in \eqref{eq: special case of 1st new bound} holds under the
requirement that the ratio $\alpha$ of the local
and total variation distances satisfies $\alpha \leq \frac{1}{N}$.
None of these conditions implies the other.
\end{remark}

\vspace*{0.1cm}
We prove in the following Theorem~\ref{theorem: 1st new bound}.
\begin{proof}
Assume without loss of generality (w.l.o.g.) that $H(X) - H(Y) \geq 0$
(note that the terms $|H(X)-H(Y)|$, $d_{\text{loc}}(X,Y)$ and $d_{\text{TV}}(X,Y)$
are invariant under a switch of $X$ and $Y$).
Let $(\hat{X}, \hat{Y})$ be the maximal coupling of $(X,Y)$ according to
the construction in the proof of Theorem~\ref{theorem: maximal coupling}. Then,
\begin{eqnarray}
&& |H(X) - H(Y)| \nonumber \\
&& = H(X) - H(Y) \nonumber \\
&& = H(\hat{X}) - H(\hat{Y}) \nonumber \\
&& = H(\hat{X}|J)-H(\hat{Y}|J) + I(\hat{X};J) - I(\hat{Y};J).
\label{eq: 1st equality in the proof of 1st new bound}
\end{eqnarray}
The conditional entropy $H(\hat{X}|J)$ satisfies
\begin{eqnarray}
&& \hspace*{-0.8cm} H(\hat{X}|J) \nonumber \\
&& \hspace*{-0.8cm} = \pr(J=0) \, H(\hat{X}|J=0) + \pr(J=1) \, H(\hat{X}|J=1) \nonumber \\
&& \hspace*{-0.8cm} \stackrel{(\text{a})}{=} d_{\text{TV}}(X,Y) \, H(V|J=0) +
\bigl(1-d_{\text{TV}}(X,Y) \bigr) \, H(U|J=1) \nonumber \\
&& \hspace*{-0.8cm} \stackrel{(\text{b})}{=} d_{\text{TV}}(X,Y) \, H(V) +
\bigl(1-d_{\text{TV}}(X,Y) \bigr) \, H(U)
\label{eq: 2nd equality in the proof of 1st new bound}
\end{eqnarray}
where equality~(a) holds since $J \sim \text{Bernoulli}(p)$
with $$p = \pr(J=1) = \pr(\hat{X} = \hat{Y}) = 1-d_{\text{TV}}(X,Y)$$
(see the proof of Theorem~\ref{theorem: maximal coupling} and the
result in Theorem~\ref{theorem: maximal coupling - second result}),
and because $\hat{X}$ is equal to $V$ or $U$ when $J$ gets that values
zero or one, respectively. Furthermore, equality~(b) holds since
$U, V, W, J$ are independent random variables (due to the construction
shown in the proof of Theorem~\ref{theorem: maximal coupling}). Similarly,
\begin{equation}
 H(\hat{Y}|J) = d_{\text{TV}}(X,Y) \, H(W) +
\bigl(1-d_{\text{TV}}(X,Y) \bigr) \, H(U).
\label{eq: 3rd equality in the proof of 1st new bound}
\end{equation}
Combining \eqref{eq: 1st equality in the proof of 1st new bound}--\eqref{eq:
3rd equality in the proof of 1st new bound} yields that
\begin{eqnarray}
|H(X)-H(Y)| = d_{\text{TV}}(X,Y) \, (H(V)-H(W)) + I(\hat{X}; J) - I(\hat{Y}; J).
\label{eq: 4th equality in the proof of 1st new bound}
\end{eqnarray}
From \eqref{eq: probability mass function of V} and
\eqref{eq: probability mass function of W}, it follows that
\begin{equation}
P_V(a) \, P_W(a) = 0, \quad \forall \, a \in \mathcal{A}
\label{eq: product of the probability mass functions of V and W}
\end{equation}
and also, for every $a \in \mathcal{A}$,
\begin{eqnarray}
&& P_V(a) + P_W(a) \nonumber \\
&& = \frac{P_X(a) + P_Y(a) - 2 \min\{P_X(a), P_Y(a)\}}{d_{\text{TV}}(X,Y)} \nonumber \\
&& = \frac{|P_X(a) - P_Y(a)|}{d_{\text{TV}}(X,Y)} \nonumber \\
&& \leq \frac{d_{\text{loc}}(X,Y)}{d_{\text{TV}}(X,Y)} \triangleq \alpha.
\label{eq: sum of the probability mass functions of V and W}
\end{eqnarray}

In the following, we derive upper bounds on $H(V)-H(W)$ and $I(\hat{X}; J) - I(\hat{Y}; J)$,
and rely on \eqref{eq: 4th equality in the proof of 1st new bound} to get an
upper bound on $|H(X)-H(Y)|$. Let
$\mathcal{A} \triangleq \{a_1, \ldots, a_M\}$, and
$$ s_i \triangleq P_V(a_i), \; \;
t_i \triangleq P_W(a_i), \quad \forall
\, i \in \{1, \ldots, M\}.$$
From \eqref{eq: product of the probability mass functions of V and W} and
\eqref{eq: sum of the probability mass functions of V and W},
\begin{equation*}
s_i t_i = 0, \quad s_i + t_i \leq \alpha, \quad \forall \, i \in \{1, \ldots, M\}
\end{equation*}
and $H(V)-H(W) = -\sum_{i=1}^M s_i \log(s_i) + \sum_{i=1}^M t_i \log(t_i).$
Hence, for fixed $\alpha$ and $M$ (since $|\mathcal{A}| = M$, then $\alpha \in [\frac{2}{M}, 1]$),
\begin{equation}
H(V)-H(W) \leq g(\alpha)
\label{eq: upper bound on the difference of the entropies of V and W}
\end{equation}
where $g(\alpha)$ is the solution of the optimization problem
\begin{eqnarray}
&& \text{maximize} \; \left(-\sum_{i=1}^M s_i \log(s_i) + \sum_{i=1}^M t_i \log(t_i) \right)
\nonumber \\
&& \text{subject to} \nonumber \\
         && \left\{ \begin{array}{c}
                     s_i, t_i \geq 0, \; s_i+t_i \leq \alpha \\
                     s_i t_i = 0, \; \; \forall \, i \in \{1, \ldots, M\} \\
                     \dsum_{i=1}^M s_i = \dsum_{i=1}^M t_i = 1
                     \end{array}
                     \right.
\label{eq: non-convex optimization problem}
\end{eqnarray}
with the $2M$ variables $s_1, t_1, \ldots s_M, t_M$. Fortunately, this {\em non-convex}
optimization problem admits a closed-form solution.
\begin{lemma}
The solution of the non-convex optimization problem in
\eqref{eq: non-convex optimization problem}, denoted
by $g(\alpha)$, is the following:
\begin{eqnarray}
g(\alpha) = \log \Bigl(M- \left\lceil\frac{1}{\alpha}\right\rceil\Bigr)
+ \alpha \, \Bigl\lfloor \frac{1}{\alpha} \Bigr\rfloor
\, \log \alpha + \left(1- \alpha \Bigl\lfloor \frac{1}{\alpha}
\Bigl \rfloor \right) \log \left( 1 - \alpha \Bigl\lfloor \frac{1}{\alpha}
\Bigr\rfloor \right)
\label{eq: closed-form expression for the solution of a non-convex optimization problem}
\end{eqnarray}
with the convention that $0 \log 0$ means~0.
\label{lemma: closed-form expression for the solution of a non-convex optimization problem}
\end{lemma}
\begin{proof}
Let's first show that the solution on the right-hand side of
\eqref{eq: closed-form expression for the solution of a non-convex optimization problem}
forms an upper bound on $g(\alpha)$, and then show that this
upper bound is tight.

For the derivation of the upper bound, note that due to the above constraints,
\begin{eqnarray}
&& 1 = \sum_{i=1}^M t_i \stackrel{(\text{a})}{\leq}
\alpha \; \bigl| \{i \in \{1, \ldots, M\}: \, t_i > 0 \} \bigr| \nonumber \\
&& \Rightarrow \; \bigl|\{ i \in \{1, \ldots, M\}: t_i > 0 \} \bigr|
\geq \frac{1}{\alpha} \nonumber \\
&& \stackrel{(\text{b})}{\Rightarrow} \; \bigl|\{ i \in \{1, \ldots, M\}: s_i > 0 \} \bigr|
\leq M - \frac{1}{\alpha} \nonumber \\[0.1cm]
&& \stackrel{(\text{c})}{\Rightarrow} \; \bigl|\{ i \in \{1, \ldots, M\}: s_i > 0 \} \bigr|
\leq M - \left\lceil\frac{1}{\alpha} \right\rceil
\label{eq: upper bound on the cardinality of the support of s_i}
\end{eqnarray}
where inequality~(a) holds since $s_i+t_i \leq \alpha$ and $s_i, t_i \geq 0$
for every $i \in \{1, \ldots, M\}$, (b) follows from the constraint
that $s_i \, t_i = 0$ for every~$i$, and (c) holds since the cardinality
of the support of $\{s_i\}$ is an integer, and
$\left\lfloor M - \frac{1}{\alpha} \right\rfloor = M - \left\lceil
\frac{1}{\alpha} \right\rceil$. Hence,
\begin{equation*}
-\sum_{i=1}^M s_i \, \log(s_i) \leq \log \Bigl(M-\left\lceil
\frac{1}{\alpha} \right\rceil\Bigr)
\end{equation*}
and the solution of the optimization problem in
\eqref{eq: non-convex optimization problem} satisfies
\begin{equation}
g(\alpha) \leq \log \Bigl(M- \Bigl\lceil \frac{1}{\alpha} \Bigr\rceil \Bigr) + f(\alpha)
\label{eq: first step in the derivation of the upper bound on g}
\end{equation}
where $f(\alpha)$ solves the optimization problem
\begin{eqnarray}
&& \text{maximize} \; \sum_{i=1}^M t_i \log(t_i)
\nonumber \\
&& \text{subject to} \nonumber \\
         && \left\{ \begin{array}{l}
                     0 \leq t_i \leq \alpha, \quad \forall \, i \in \{1, \ldots, M\} \\
                     \dsum_{i=1}^M t_i = 1
                     \end{array}
                     \right.
\label{eq: a more simple optimization problem}
\end{eqnarray}
with the $M$ optimization variables $t_1, \ldots, t_M$. Note that the objective
function in \eqref{eq: a more simple optimization problem} is convex, and the feasible
set is a bounded polyhedron. Furthermore, the maximum of a convex function over a
bounded polyhedron is attained at one of its vertices (see, e.g.,
\cite[Corollary~32.3.3]{Rockafellar}; this property follows from the convex-hull
description of a bounded polyhedron and Jensen's inequality). Since
the objective function and the feasible set in \eqref{eq: a more simple optimization problem}
are invariant to a permutation of the variables $t_1, \ldots, t_M$, then an optimal point
is given by
\begin{eqnarray}
&& t_1 = \ldots = t_l = \alpha,
\quad l \triangleq \Bigl\lfloor \frac{1}{\alpha} \Bigr\rfloor \nonumber \\
&& t_{l+1} = 1 - \alpha \, \Bigl\lfloor \frac{1}{\alpha} \Bigr\rfloor,  \quad
t_{l+2} = \ldots = t_M = 0 \nonumber
\end{eqnarray}
where $l \leq \frac{M}{2}$ (since $\alpha \in [\frac{2}{M}, 1]$); as requested,
$t_i \in [0, \alpha]$ for $i \in \{1, \ldots, M\}$.
This implies that the solution of the optimization problem in
\eqref{eq: a more simple optimization problem} is given by
\begin{equation}
\hspace*{-0.2cm} f(\alpha) = \alpha \, \Bigl\lfloor \frac{1}{\alpha} \Bigr\rfloor
\, \log \alpha + \left(1- \alpha \Bigl\lfloor \frac{1}{\alpha}
\Bigl \rfloor \right) \log \left( 1 - \alpha \Bigl\lfloor \frac{1}{\alpha}
\Bigr\rfloor \right).
\label{eq: exact closed-form expression for f}
\end{equation}
From \eqref{eq: first step in the derivation of the upper bound on g} and
\eqref{eq: exact closed-form expression for f}, it follows that the right-hand side
of \eqref{eq: closed-form expression for the solution of a non-convex optimization problem}
forms an upper bound on $g(\alpha)$. It remains to show that this bound is tight. To
this end, we separate into the following two cases:

{\em Case 1}: Suppose that $N \triangleq \frac{1}{\alpha}$ is an integer. In this case,
the upper bound on $g(\alpha)$ (see \eqref{eq: first step in the derivation of the upper bound on g}
and \eqref{eq: exact closed-form expression for f})
gets the simplified form
$$g(\alpha) \leq \log \Bigl(M - \frac{1}{\alpha}\Bigr) + \log \alpha = \log(M \alpha-1).$$
This upper bound on $g(\alpha)$ is achieved by the point $(s_1, t_1, \ldots, s_M, t_M)$
where
\begin{eqnarray*}
&& t_1 = \ldots = t_N = \alpha, \quad t_{N+1} = \ldots = t_M = 0 \\
&& s_1 = \ldots = s_N = 0, \quad s_{N+1} = \ldots = s_M = \frac{1}{M-N} \, .
\end{eqnarray*}
Note that this point is included in the feasible set of the optimization problem in
\eqref{eq: non-convex optimization problem} since
$ \frac{1}{M-N} = \frac{\alpha}{M \alpha - 1} \leq \alpha$ where the last inequality
holds because $\alpha \in [\frac{2}{M}, 1]$. The value of the objective
function in \eqref{eq: non-convex optimization problem} at this point is
equal to
\begin{eqnarray*}
&& -\sum_{i=1}^M s_i \log(s_i) + \sum_{i=1}^M t_i \log(t_i) \\
&& = \log \Bigl(M - \frac{1}{\alpha} \Bigr) + \log \alpha = \log(M \alpha-1)
\end{eqnarray*}
so this upper bound on $g(\alpha)$ is tight if $\frac{1}{\alpha}$ is an integer.

{\em Case 2}: Suppose that $\frac{1}{\alpha}$ is not an integer. In this case, let
$l \triangleq \left\lfloor \frac{1}{\alpha} \right\rfloor$ so
$l+1 = \Bigl \lceil \frac{1}{\alpha} \Bigr \rceil$, and consider the
$(2M)$-dimensional vector $(s_1, t_1, \ldots, s_M, t_M)$ where
\begin{eqnarray}
&& t_1 = \ldots = t_l = \alpha, \quad t_{l+1} = 1 - \alpha \Bigl\lfloor \frac{1}{\alpha}
\Bigr\rfloor \nonumber \\
&& t_{l+2} = \ldots = t_M = 0 \nonumber \\
&& s_1 = \ldots = s_{l+1} = 0  \label{eq: selected point to show tightness of the bound in 2nd case} \\
&& s_{l+2} = \ldots = s_M = \frac{1}{M-l-1} = \frac{1}{M - \Bigl \lceil \frac{1}{\alpha} \Bigr \rceil} \, . \nonumber
\end{eqnarray}
To verify that it is included in the feasible set of
\eqref{eq: non-convex optimization problem}, note that due to the constraints of
this optimization problem
\begin{eqnarray*}
&& 1 = \sum_{i=1}^M s_i {\leq} \alpha \; \bigl|
\{i \in \{1, \ldots, M\}: \, s_i > 0 \} \bigr| \\
&& \Rightarrow \; \bigl|\{ i \in \{1, \ldots, M\}: s_i > 0 \} \bigr|
\geq \frac{1}{\alpha} \nonumber \\
&& \Rightarrow \; \bigl|\{ i \in \{1, \ldots, M\}: s_i > 0 \} \bigr|
\geq \left\lceil \frac{1}{\alpha} \right\rceil
\end{eqnarray*}
and, by combining it with \eqref{eq: upper bound on the cardinality of the support of s_i},
it follows that
$$\left\lceil \frac{1}{\alpha} \right\rceil \leq
\bigl|\{ i \in \{1, \ldots, M\}: s_i > 0 \} \bigr|
\leq M - \left\lceil\frac{1}{\alpha} \right\rceil$$
so $\left\lceil\frac{1}{\alpha} \right\rceil \leq \frac{M}{2}$.
This implies that for $j \in \{l+2, \ldots, M\}$ (note also that
$\alpha \in [\frac{2}{M}, 1]$)
\begin{eqnarray*}
&& \hspace*{-0.3cm} s_j = \frac{1}{M-l-1} \\
&& = \frac{1}{M - \left\lceil \frac{1}{\alpha} \right\rceil} \\
&& \leq \frac{2}{M} \\
&& \leq \alpha
\end{eqnarray*}
and $t_{l+1} = 1 - \alpha \left\lfloor \frac{1}{\alpha} \right\rfloor \leq \alpha$,
so the vector is indeed included in the feasible set of
\eqref{eq: non-convex optimization problem}. The value of the objective
function in \eqref{eq: non-convex optimization problem} at the selected point
in \eqref{eq: selected point to show tightness of the bound in 2nd case} is equal to
\begin{eqnarray*}
&& \hspace*{-0.3cm} -\sum_{i=1}^M s_i \, \log(s_i) + \sum_{i=1}^M t_i \log(t_i) \\
&& \hspace*{-0.3cm} = \log \Bigl(M- \left\lceil\frac{1}{\alpha}\right\rceil\Bigr)
+ \alpha \, \Bigl\lfloor \frac{1}{\alpha} \Bigr\rfloor
\, \log \alpha + \left(1- \alpha \Bigl\lfloor \frac{1}{\alpha}
\Bigl \rfloor \right) \log \left( 1 - \alpha \Bigl\lfloor \frac{1}{\alpha}
\Bigr\rfloor \right) \\
&& \hspace*{-0.3cm} = g(\alpha)
\end{eqnarray*}
so the upper bound on $g(\alpha)$ from
\eqref{eq: first step in the derivation of the upper bound on g} and
\eqref{eq: exact closed-form expression for f} is tight, and this completes the proof of
Lemma~\ref{lemma: closed-form expression for the solution of a non-convex optimization problem}.
\end{proof}

\vspace*{0.1cm}
\begin{corollary}
The solution of the non-convex optimization problem in \eqref{eq: non-convex optimization problem}
satisfies the inequality
\begin{equation*}
g(\alpha) \leq \log(M \alpha - 1)
\end{equation*}
and this bound is tight if and only if $\frac{1}{\alpha}$ is an integer.
\label{corollary: simple bound on g}
\end{corollary}
\begin{proof}
From Lemma~\ref{lemma: closed-form expression for the solution of a non-convex optimization problem}
(see Eq.~\eqref{eq: closed-form expression for the solution of a non-convex optimization problem}), it follows that
\begin{eqnarray*}
&& g(\alpha) \nonumber \\
&& \leq \log \Bigl( M - \frac{1}{\alpha} \Bigr)
+ \alpha \, \Bigl\lfloor \frac{1}{\alpha} \Bigr\rfloor
\, \log \alpha + \left(1- \alpha \Bigl\lfloor \frac{1}{\alpha}
\Bigl \rfloor \right) \log \left( 1 - \alpha \Bigl( \frac{1}{\alpha} - 1 \Bigr) \right)
\nonumber \\
&& = \log \Bigl( M - \frac{1}{\alpha} \Bigr) + \log(\alpha) \nonumber \\
&& = \log(M \alpha -1)
\end{eqnarray*}
and the above inequality turns to be an equality if and only if $\frac{1}{\alpha}$
is an integer.
\end{proof}

By combining
\eqref{eq: upper bound on the difference of the entropies of V and W}
and Corollary~\ref{corollary: simple bound on g}, it follows that
\begin{equation*}
H(V) - H(W) \leq \log (M \alpha-1)
\end{equation*}
and therefore from \eqref{eq: 4th equality in the proof of 1st new bound}
\begin{eqnarray}
|H(X)-H(Y)| \leq d_{\text{TV}}(X,Y) \, \log(M \alpha - 1) + I(J; \hat{X}) - I(J; \hat{Y}).
\label{eq: one step before the derivation of 1st bound}
\end{eqnarray}
Finally, the bound in \eqref{eq: 1st new bound} follows from the inequality
\begin{equation}
I(J; \hat{X}) - I(J; \hat{Y}) \leq H(J) = h\bigl(d_{\text{TV}}(X,Y)\bigr).
\label{eq: upper bound on the difference of two MI}
\end{equation}

We move to derive a refinement of the bound in \eqref{eq: 1st new bound}
when $\frac{1}{2} \leq \frac{P_X}{P_Y} \leq 2$. In this case, the starting point is
the inequality in \eqref{eq: one step before the derivation of 1st bound} where it
is aimed to improve the upper bound in
\eqref{eq: upper bound on the difference of two MI}. To this end,
\begin{eqnarray}
&& I(J; \hat{X}) - I(J; \hat{Y}) \nonumber \\
&& = H(J|\hat{Y}) - H(J|\hat{X}) \nonumber \\
&& \leq H(J) -  H(J|\hat{X}) \nonumber \\
&& = h\bigl(d_{\text{TV}}(X,Y)\bigr) - H(J|\hat{X})
\label{eq: refined upper bound on the difference of two MI}
\end{eqnarray}
and, from \cite[Theorem~11]{conditional_entropy_and_error_probability_IT2010},
\begin{equation}
H(J|\hat{X}) \geq 2 \log 2 \; \pr\bigl(J \neq J_{\text{MAP}}(\hat{X})\bigr)
\label{eq: lower bound on the conditional entropy}
\end{equation}
where $J_{\text{MAP}}(\hat{X})$ is the maximum a-posteriori (MAP) estimator
of $J$ based on $\hat{X}$ (note that the minimum on the
left-hand side of \cite[Eq.~(110)]{conditional_entropy_and_error_probability_IT2010}
is achieved by the MAP estimator). In the following, the estimator
$J_{\text{MAP}}(\hat{X})$ on the right-hand side of
\eqref{eq: lower bound on the conditional entropy} is calculated.
\begin{enumerate}
\item If $\hat{X} \notin \text{supp}(P_V)$ then a.s. $J=1$ (otherwise, $J=0$
and $\hat{X}=V$, so $\hat{X} \in \text{supp}(P_V)$ a.s.). Hence,
$$\hat{X} \notin \text{supp}(P_V) \; \Rightarrow \; J_{\text{MAP}}(\hat{X})=1.$$
From \eqref{eq: probability mass function of V}, it follows that
$\hat{X} \notin \text{supp}(P_V)$
if and only if $P_X(\hat{X}) \leq P_Y(\hat{X})$.
\item If $\hat{X} \in \text{supp}(P_V)$ then, from \eqref{eq: probability mass function of V},
$P_X(\hat{X}) > P_Y(\hat{X})$. Hence, from
\eqref{eq: probability mass function of U} and \eqref{eq: probability mass function of V}
with $p = 1-d_{\text{TV}}(X,Y)$,
\begin{eqnarray*}
&& P_U(\hat{X}) = \frac{P_Y(\hat{X})}{1-d_{\text{TV}}(X,Y)} \\
&& P_V(\hat{X}) = \frac{P_X(\hat{X})-P_Y(\hat{X})}{d_{\text{TV}}(X,Y)}.
\end{eqnarray*}
Since $U, V, J$ are independent, then from \eqref{eq: probability mass function of J}
\begin{eqnarray*}
&& \hspace*{-1cm} \pr(J=1, \hat{X}) = \pr(J=1) \, P_U(\hat{X}) = P_Y(\hat{X}) \\
&& \hspace*{-1cm} \pr(J=0, \hat{X}) = \pr(J=0) \, P_V(\hat{X}) = P_X(\hat{X}) - P_Y(\hat{X})
\end{eqnarray*}
so, if $\hat{X} \in \text{supp}(P_V)$, then
\begin{eqnarray*}
J_{\text{MAP}}(\hat{X}) = \left\{ \begin{array}{ll}
                          1
                          & \mbox{if $\frac{P_X(\hat{X})}{2} \leq P_Y(\hat{X}) < P_X(\hat{X}) $} \\
                          0  & \mbox{if $P_Y(\hat{X}) < \frac{P_X(\hat{X})}{2}$}.
                          \end{array}
                          \right.
\end{eqnarray*}
\end{enumerate}
To conclude, the MAP estimator of $J$ that is based on the observation $\hat{X}$
is given by
\begin{eqnarray*}
J_{\text{MAP}}(\hat{X}) = \left\{ \begin{array}{ll}
                          1
                          & \mbox{if $\frac{P_X(\hat{X})}{2} \leq P_Y(\hat{X})$} \\
                          0  & \mbox{if $P_Y(\hat{X}) < \frac{P_X(\hat{X})}{2}$}.
                          \end{array}
                          \right.
\label{eq: MAP estimator of J based on X hat}
\end{eqnarray*}
It therefore implies that if $\frac{P_Y}{P_X} \geq \frac{1}{2}$ whenever $P_X>0$,
then $J_{\text{MAP}}(\hat{X})=1$ independently of $\hat{X}$, so in this case
$$ \pr\bigl(J \neq J_{\text{MAP}}(\hat{X})\bigr) = \pr(J=0) = d_{\text{TV}}(X,Y).$$
Hence, from \eqref{eq: refined upper bound on the difference of two MI},
\eqref{eq: lower bound on the conditional entropy} and the last equality,
if $\frac{P_Y}{P_X} \geq \frac{1}{2}$ whenever $P_X>0$ then
\begin{equation*}
\hspace*{-0.2cm} I(J; \hat{X}) - I(J; \hat{Y}) \leq h\bigl(d_{\text{TV}}(X,Y)\bigr)
- 2 \log 2 \cdot d_{\text{TV}}(X,Y).
\end{equation*}
A combination of the last inequality with \eqref{eq: one step before the derivation of 1st bound}
finally gives the refined bound in \eqref{eq: possible refinement of 1st new bound}.
Since it was assumed at the beginning of the proof that $H(X) \geq H(Y)$
while it is not necessarily known in advance which entropy is larger, the requirement on
$\frac{P_Y}{P_X}$ can be symmetrized by requiring that
$\frac{1}{2} \leq \frac{P_X}{P_Y} \leq 2$ whenever $P_X, P_Y > 0$.
This completes the proof of Theorem~\ref{theorem: 1st new bound}.
\end{proof}

\vspace*{0.2cm}
\begin{corollary}
Let $X$ and $Y$ be two discrete random variables that take values
in a finite set $\mathcal{A}$, and let $|\mathcal{A}|=M$. Assume
that for some positive constants $\varepsilon_1, \varepsilon_2$
\begin{eqnarray}
&& d_{\text{TV}}(X,Y) \leq \varepsilon_1 \leq 1-\frac{1}{M \varepsilon_2} \, ,
\label{eq: condition 1 in 1st corollary} \\
&& \frac{d_{\text{loc}}(X,Y)}{d_{\text{TV}}(X,Y)} \leq \varepsilon_2 \leq 1.
\label{eq: condition 2 in 1st corollary}
\end{eqnarray}
Then,
\begin{equation}
|H(X)-H(Y)| \leq \varepsilon_1 \, \log(M \varepsilon_2-1) + h(\varepsilon_1).
\label{eq: entropy bound in the first new corollary}
\end{equation}
\label{corollary: first new corollary}
\end{corollary}
\vspace*{-0.4cm}
\begin{proof}
From \eqref{eq: 1st new bound}, \eqref{eq: alpha}, \eqref{eq: condition 2 in 1st corollary},
and since $\alpha \leq \varepsilon_2$
\begin{equation*}
|H(X)-H(Y)| \leq d_{\text{TV}}(X,Y) \, \log(M \varepsilon_2-1) + h\bigl(d_{\text{TV}}(X,Y)\bigr).
\end{equation*}
The function $q(\varepsilon) \triangleq \varepsilon c + h(\varepsilon)$ is
monotonic increasing over the interval $\Bigl[0, \frac{e^c}{1+e^c}\Bigr]$
($q'(\varepsilon) = c + \log\left(\frac{1-\varepsilon}{\varepsilon}\right) > 0$
if and only if $0 < \varepsilon < \frac{e^c}{1+e^c}$). Referring to the right-hand
side of the above inequality, let $c \triangleq \log(M \varepsilon_2-1)$, so
$\frac{e^c}{1+e^c} = 1-\frac{1}{M \varepsilon_2}$. Hence, if the conditions in
\eqref{eq: condition 1 in 1st corollary} and \eqref{eq: condition 2 in 1st corollary}
are satisfied then the inequality in \eqref{eq: entropy bound in the first new corollary}
holds.
\end{proof}

\vspace*{0.1cm}
\begin{remark}
By considering the pair of probability mass functions $P_{X,Y}$ and
$P_X \times P_Y$ (without abuse of notation, let
$H(P_X) \triangleq H(X)$), then
\begin{eqnarray*}
&& H(P_X \times P_Y) - H(P_{X,Y}) \\
&& = H(X)+H(Y)-H(X,Y) \\
&& = I(X;Y).
\end{eqnarray*}
Hence, Theorem~\ref{theorem: 1st new bound} and Corollary~\ref{corollary: first new corollary}
provide bounds on the mutual information between two discrete random variables of finite
support, where these bounds are expressed in terms of the local and total variation distances
between the joint distribution of $(X,Y)$ and the product of its marginal distributions. The
specialization of Theorem~\ref{theorem: 1st new bound} to this setting tightens the bound in
\cite[Theorem~1]{Zhang_IT2007}, and the former bound is particularized to the latter known bound
in the case where the local and total variation distances are equal (which is the extreme case).
\end{remark}

\vspace*{0.1cm}
\begin{remark}
The bound in \cite[Theorem~1]{Zhang_IT2007} was improved in \cite[Proposition~1]{Prelov_PPI2008}
without any further assumptions. It is noted that by introducing the additional requirement
where there exists some constant $\varepsilon_2 \in [0,1]$ such that for every $y \in \mathcal{Y}$
$$\frac{d_{\text{loc}}(P_X, P_{X| Y=y})}{d_{\text{TV}}(P_X, P_{X| Y=y})} \leq \varepsilon_2$$
then it enables to refine the bound in \cite[Proposition~1]{Prelov_PPI2008}.
This follows by combining the proof of
\cite[Proposition~1]{Prelov_PPI2008} with \eqref{eq: entropy bound in the first new corollary}
(see Corollary~\ref{corollary: first new corollary}) where
Eq.~\eqref{eq: entropy bound in the first new corollary}
replaces the use of \cite[Eq.~(4)]{Zhang_IT2007} in \cite[Eq.~(35)]{Prelov_PPI2008}.
The same thing also applies to \cite[Proposition~2]{Prelov_PPI2009}, referring to its
proof in \cite[p.~305]{Prelov_PPI2009}.
\end{remark}

\vspace*{0.2cm}
For countably infinite alphabets, just knowing the total variation distance
between two distributions does not imply anything about the
difference of entropies (i.e., one has discontinuity of the entropy). The
following theorem shows that is one of the distributions is finitely supported,
and some knowledge of the tail behavior of the other distribution is available,
then having bounds on the local and total variation distances allows one to
bound the difference of the entropies even in this case.
\begin{theorem}
Let $\mathcal{A} = \{a_1, a_2, \ldots \}$ be a countable infinite set.
Let $X$ and $Y$ be discrete random variables where $X$ takes values
in the set $\mathcal{X} = \{a_1, \ldots, a_m\}$ for some $m \in \naturals$,
and $Y$ takes values in the set $\mathcal{A}$. Assume that for some
$\eta_1, \eta_2, \eta_3 > 0$, the local and total variation distances between
$X$ and $Y$ satisfy
\begin{equation}
\eta_2 \leq d_{\text{TV}}(X,Y) \leq \eta_1, \quad d_{\text{loc}}(X,Y) \leq \eta_3
\label{eq: conditions on TV and local distances}
\end{equation}
where $\eta_3 \leq \eta_2 \leq \eta_1 < 1$. Let $M$ be an integer such that
\begin{equation}
\sum_{i=M}^{\infty} P_Y(a_i) \leq \eta_3, \quad
M \geq \max\biggl\{m+1, \frac{\eta_2}{(1-\eta_1)\eta_3}\biggr\}
\label{eq: conditions on M in the second new theorem}
\end{equation}
and let $\eta_4 > 0$ satisfy
\begin{equation}
-\sum_{i=M}^{\infty} P_Y(a_i) \, \log P_Y(a_i) \leq \eta_4.
\end{equation}
Then, the following inequality holds:
\begin{equation}
|H(X)-H(Y)| \leq \eta_1 \, \log \biggl(\frac{M \eta_3}{\eta_2} - 1 \biggr) + h(\eta_1) + \eta_4.
\label{eq: bound is 2nd theorem}
\end{equation}
\label{theorem: second new theorem for the case of an infinite countable set A}
\end{theorem}
\begin{remark}
The inequality $\eta_3 \leq \eta_2 \leq \eta_1 < 1$ (after
\eqref{eq: conditions on TV and local distances}) is easily satisfied since
$d_{\text{loc}}(X,Y) \leq d_{\text{TV}}(X,Y) \leq 1$; so, if $d_{\text{TV}}(X,Y)<1$
then it is possible to choose $\eta_1$, $\eta_2$ and $\eta_3$ that satisfy this inequality.
\end{remark}
\begin{proof}
Let $\widetilde{Y}$ be a random variable that is defined to be
equal to $Y$ if $Y \in \{a_1, \ldots, a_{M-1}\}$, and it is
set to be equal to $a_M$ if $Y=a_i$ for some $i \geq M$. Hence, the probability
mass function of $\widetilde{Y}$ is related to that of $Y$ as follows:
\begin{equation}
P_{\widetilde{Y}}(a_i) = \left\{ \begin{array}{ll}
                          P_Y(a_i)  & \mbox{if $i \in \{1, \ldots, M-1\}$} \\[0.1cm]
                          \sum_{j=M}^{\infty} P_Y(a_j)  & \mbox{if $i=M$.}
                          \end{array}
                          \right.
\label{eq: probability mass function of Y tilde}
\end{equation}
Since $P_X(a_i) = 0$ for every $i > m$ and also $M \geq m+1$ (see the second inequality
in \eqref{eq: conditions on M in the second new theorem}), then it follows from
\eqref{eq: probability mass function of Y tilde} that
\begin{eqnarray}
&& \hspace*{-0.7cm} d_{\text{TV}}(X, \widetilde{Y}) \nonumber \\
&& \hspace*{-0.7cm} = \frac{1}{2} \sum_{i=1}^m |P_X(a_i) - P_{\widetilde{Y}}(a_i)|
+ \frac{1}{2} \sum_{i=m+1}^M P_{\widetilde{Y}}(a_i) \nonumber \\
&& \hspace*{-0.7cm} = \frac{1}{2} \sum_{i=1}^m |P_X(a_i) - P_Y(a_i)| +
\frac{1}{2} \sum_{i=m+1}^{\infty} P_Y(a_i) \nonumber \\
&& \hspace*{-0.7cm} = d_{\text{TV}}(X, Y).
\label{eq: the two total variation distances are equal}
\end{eqnarray}
Hence, $X$ and $\widetilde{Y}$ are discrete random variables that take values
in the set $\{a_1, \ldots, a_M\}$ (note that it includes the set $\mathcal{X}$),
and from \eqref{eq: conditions on TV and local distances} and
\eqref{eq: the two total variation distances are equal}
\begin{equation}
0< \eta_2 \leq d_{\text{TV}}(X, \widetilde{Y}) \leq \eta_1.
\label{eq: bounds on TV distance between X and Y tilde}
\end{equation}
Furthermore, the local distance between $X$ and $\widetilde{Y}$ satisfies
\begin{eqnarray}
&& d_{\text{loc}}(X, \widetilde{Y}) \nonumber \\
&& = \max_{i \in \{1, \ldots, M\}} |P_X(a_i) - P_{\widetilde{Y}}(a_i)| \nonumber \\
&& \stackrel{(\text{a})}{=} \max\biggl\{ \max_{i \in \{1, \ldots, M-1\}} |P_X(a_i) - P_Y(a_i)| \, ,
\sum_{i=M}^{\infty} P_Y(a_i) \biggr\} \nonumber \\
&& \stackrel{(\text{b})}{\leq} \max \{d_{\text{loc}}(X,Y), \eta_3 \} \nonumber \\
&& \stackrel{(\text{c})}{=} \eta_3
\label{eq: local distance between X and tilde Y}
\end{eqnarray}
where (a), (b) and (c) above follow from the equality in
\eqref{eq: probability mass function of Y tilde} (note also that $m \leq M-1$),
the first inequality in \eqref{eq: conditions on M in the second new theorem} and
the second inequality in \eqref{eq: conditions on TV and local distances}, respectively.
From \eqref{eq: bounds on TV distance between X and Y tilde} and
\eqref{eq: local distance between X and tilde Y}
\begin{eqnarray}
&& d_{\text{TV}}(X, \widetilde{Y}) \leq \eta_1 \triangleq \varepsilon_1
\label{eq: epsilon_1} \\[0.1cm]
&& \frac{d_{\text{loc}}(X, \widetilde{Y})}{d_{\text{TV}}(X, \widetilde{Y})} \leq \frac{\eta_3}{\eta_2} \triangleq \varepsilon_2
\label{eq: epsilon_2}
\end{eqnarray}
where $0 < \varepsilon_1 < 1$ and $0 < \varepsilon_2 \leq 1$ (since, by assumption,
$0< \eta_3 \leq \eta_2 \leq \eta_1 < 1$).
The integer $M$ is set to satisfy the inequality $M \geq \frac{\eta_2}{\eta_3 (1-\eta_1)}$
(see \eqref{eq: conditions on M in the second new theorem}),
so from \eqref{eq: epsilon_1} and \eqref{eq: epsilon_2}
$$\varepsilon_1 \leq 1-\frac{1}{M \varepsilon_2}.$$
Hence, it follows from Theorem~\ref{theorem: 1st new bound} that
\begin{equation}
|H(X) - H(\widetilde{Y})| \leq \eta_1 \, \log\Bigl(\frac{M \eta_3}{\eta_2} - 1\Bigr) + h(\eta_1).
\label{eq: bound on the entropy difference in terms of eta1, eta2, eta3}
\end{equation}
Since $\widetilde{Y}$ is a deterministic function of $Y$ then
$H(Y) \geq H(\widetilde{Y})$, and from \eqref{eq: probability mass function of Y tilde}
\begin{eqnarray}
&& | H(\widetilde{Y}) - H(Y) | \nonumber \\
&& = H(Y) - H(\widetilde{Y}) \nonumber \\[0.1cm]
&& = - \sum_{i=M}^{\infty} P_Y (a_i) \, \log P_Y(a_i)
+ \left( \sum_{i=M}^{\infty} P_Y(a_i) \right)
\log \left( \sum_{i=M}^{\infty} P_Y(a_i) \right) \nonumber \\
&& \leq - \sum_{i=M}^{\infty} P_Y (a_i) \, \log P_Y(a_i) \leq \eta_4.
\label{eq: bound on the entropy difference of Y and Y tilde}
\end{eqnarray}
Finally, the bound in \eqref{eq: bound is 2nd theorem} follows from
\eqref{eq: bound on the entropy difference in terms of eta1, eta2, eta3},
\eqref{eq: bound on the entropy difference of Y and Y tilde} and the triangle inequality.
\end{proof}

\begin{corollary}
In the setting of $X$ and $Y$ in
Theorem~\ref{theorem: second new theorem for the case of an infinite countable set A},
assume that $d_{\text{TV}}(X,Y) \leq \eta$ for some $\eta \in (0,1)$. Let
$M \triangleq \max\Bigl\{m+1, \frac{1}{1-\eta}\Bigr\}$,
and assume that for some $\mu > 0$
\begin{equation*}
-\sum_{i=M}^{\infty} P_Y(a_i) \, \log P_Y(a_i) \leq \mu
\end{equation*}
then
$|H(X)-H(Y)| \leq \eta \log(M-1) + h(\eta) + \mu.$
\label{corollary: 2nd new corollary}
\end{corollary}
\begin{proof}
This follows from
Theorem~\ref{theorem: second new theorem for the case of an infinite countable set A}
by setting $\eta_2 = \eta_3 = d_{\text{loc}}(X,Y)$ (note that
$d_{\text{loc}}(X,Y) \leq d_{\text{TV}}(X,Y)$), and then
$\eta_1$ and $\eta_4$ are replaced by $\eta$ and $\mu$, respectively.
\end{proof}

\begin{remark}
The result in Corollary~\ref{corollary: 2nd new corollary} can be obtained
by a simplification of the proof of
Theorem~\ref{theorem: second new theorem for the case of an infinite countable set A}
where \eqref{eq: bound on the entropy difference in terms of eta1, eta2, eta3}
is replaced by the bound in \cite[Eq.~(4)]{Zhang_IT2007} (see \eqref{eq: Zhang's bound}),
without the refinement which takes the local distance into consideration.
\end{remark}

\section{Examples}
\label{section: Examples}
In the following, we exemplify the use of the new bounds in
Section~\ref{section: Refined Bounds on the Entropy of Discrete Random Variables via Coupling},
and also compare them with some previously known bounds.

\begin{example}
Let $X$ be a discrete random variable that gets values in
the set $\mathcal{A} = \{a_1, \ldots, a_M\}$.
Let's express its arbitrary probability mass function in the form
\begin{equation}
P_X(a_i) = \frac{1+u_i \xi_i}{M} \quad \forall \, i \in \{1, \ldots, M\}
\label{eq: probability mass function of first example}
\end{equation}
where
\begin{eqnarray*}
&& u_i \in \{-1, 1\}, \; \xi_i \geq 0, \\
&& 0 \leq 1 + u_i \xi_i \leq M, \quad \forall \, i \in \{1, \ldots, M\}
\end{eqnarray*}
and $$\sum_{i=1}^M u_i \xi_i = 0$$ where the latter equality is equivalent to
$\sum_{i=1}^M P_X(a_i)=1$.

In the following, we derive a lower bound on the entropy $H(X)$. Let
$Y$ be a random variable that takes the values
from $\mathcal{A}$ with equal probability, so $H(Y) = \log M$. The local and
total variation distances between $X$ and $Y$ are equal to
\begin{eqnarray*}
&& d_{\text{TV}}(X,Y) = \frac{1}{2M} \sum_{i=1}^M \xi_i = \frac{\xi_{\text{avg}}^{(M)}}{2} \\
&& d_{\text{loc}}(X,Y) = \frac{1}{M} \, \max_{1 \leq i \leq M} \xi_i = \frac{\xi_{\max}^{(M)}}{M}
\end{eqnarray*}
where $\xi_{\text{avg}}^{(M)}$ and $\xi_{\max}^{(M)}$ denote the average and maximal values of
$\{\xi_i\}_{i=1}^M$, respectively. From \eqref{eq: alpha}
$$\alpha_M \triangleq \frac{d_{\text{loc}}(X,Y)}{d_{\text{TV}}(X,Y)} =
\frac{2 \xi_{\max}^{(M)}}{M \xi_{\text{avg}}^{(M)}}$$
so $$\alpha_M = \frac{2K_M}{M}$$ where
\begin{equation}
K_M \triangleq \frac{\xi_{\max}^{(M)}}{\xi_{\text{avg}}^{(M)}}.
\label{eq: K_M}
\end{equation}
From \eqref{eq: 1st new bound} (where also $H(Y) = \log M \geq H(X)$), it follows that
\begin{equation*}
\log M - \frac{\xi_{\text{avg}}^{(M)} \; \log(2K_M-1)}{2} - h\biggl(\frac{\xi_{\text{avg}}^{(M)}}{2}\biggr) \leq H(X) \leq \log M
\end{equation*}
and, since the binary entropy function is bounded between 0 and $\log 2$, the above
inequality can be loosened to
\begin{equation}
1 - \frac{\xi_{\text{avg}}^{(M)}}{2} \, \frac{\log(2K_M-1)}{\log M} - \frac{\log 2}{\log M}
\leq \frac{H(X)}{\log M} \leq 1
\label{eq: bounds on the normalized entropy of X from the 1st new theorem}
\end{equation}
which implies (since $K_M \geq 1$) that
\begin{equation}
\lim_{M \rightarrow \infty} \frac{\xi_{\text{avg}}^{(M)} \, \log K_M}{\log M} = 0
\; \Rightarrow \; \lim_{M \rightarrow \infty} \frac{H(X)}{\log M} = 1.
\label{eq: condition from the 1st new theorem}
\end{equation}
For comparison, the bound in
Theorem~\ref{theorem: known bound on the entropy difference in terms of total variation distance}
gives that
\begin{equation}
1 - \frac{\xi_{\text{avg}}^{(M)}}{2} \cdot \frac{\log(M-1)}{\log M} -
\frac{1}{\log M} \cdot h\biggl(\frac{\xi_{\text{avg}}^{(M)}}{2}\biggr)
\leq \frac{H(X)}{\log M} \leq 1
\label{eq: bounds on the entropy from the known theorem}
\end{equation}
which implies that
\begin{equation}
\lim_{M \rightarrow \infty} \xi_{\text{avg}}^{(M)} = 0
\; \Rightarrow \;
\lim_{M \rightarrow \infty} \frac{H(X)}{\log M} = 1.
\label{eq: stronger condition from the known theorem}
\end{equation}
The latter condition in \eqref{eq: stronger condition from the known theorem}
is stronger than \eqref{eq: condition from the 1st new theorem}. To
see this, note that $1 \leq K_M \leq \frac{M}{2}$ (since
$\frac{2}{M} \leq \frac{d_{\text{loc}}(X,Y)}{d_{\text{TV}}(X,Y)} \leq 1$).
On the other hand, as a concrete example for the case where the condition
in \eqref{eq: condition from the 1st new theorem} holds whereas the condition
in \eqref{eq: stronger condition from the known theorem} does not hold,
let $M$ be an arbitrary even number, and
$$u_i = (-1)^i, \; \xi_i = \beta \in (0,1], \quad i \in \{1, \ldots, M\}$$
where, indeed, $\sum_{i=1}^M u_i \xi_i = \beta \sum_{i=1}^M (-1)^i = 0$.
In this case, $P_X(a_i) = \frac{1-\beta}{M}$ for odd numbers
$i \in \{1, \ldots, M\}$, and $P_X(a_i) = \frac{1+\beta}{M}$ for
even numbers $i$. Furthermore, in this case $K_M=1$ for every even $M$, so
the condition in \eqref{eq: condition from the 1st new theorem} holds
by letting the even number $M$ tend to infinity.
On the other hand, the condition in \eqref{eq: stronger condition from the known theorem}
is not satisfied since
$\lim_{M \rightarrow \infty} \xi_{\text{avg}}^{(M)} = \beta > 0.$
The upper and lower bounds in \eqref{eq: bounds on the entropy from the known theorem}
tend to $1$ and $1-\frac{\beta}{2}$, respectively, so the gap between
these asymptotic bounds is increased linearly with $\beta$.
Therefore, Theorem~\ref{theorem: 1st new bound} gives a simple
lower bound on the entropy $H(X)$ in terms of the average
and maximal values of $\{\xi_i\}_{i=1}^M$, which improves the lower
bound on the entropy that follows from the known bound in
Theorem~\ref{theorem: known bound on the entropy difference in terms of total variation distance}
(see \eqref{eq: bounds on the entropy from the known theorem}).

For comparison, the bound in \cite[Theorem~7]{entropy_difference_and_variational_distance_IT2010}
is also applied to this example. In this case, since $P_X, P_Y \leq \frac{1+\xi_{\max}}{M}$
then $P_X$ and $P_Y$ are less than or equal to $\frac{1}{N_M}$ with
$N_M \triangleq \Bigl\lfloor \frac{M}{1+\xi_{\max}^{(M)}} \Bigr\rfloor$. Similarly to the above analysis,
it is easy to verify from
\cite[Theorem~7]{entropy_difference_and_variational_distance_IT2010} that
\begin{equation}
\lim_{M \rightarrow \infty} \frac{\xi_{\text{avg}}^{(M)}
\log \bigl(\xi_{\max}^{(M)} \bigr)}{\log M} = 0
\; \Rightarrow \;
\lim_{M \rightarrow \infty} \frac{H(X)}{\log M} = 1.
\label{eq: condition related to the improved known bound}
\end{equation}
Since
$$ \frac{\xi_{\text{avg}}^{(M)}
\log \bigl(\xi_{\max}^{(M)} \bigr)}{\log M} \geq \frac{\xi_{\text{avg}}^{(M)}
\log \bigl(\xi_{\text{avg}}^{(M)} \bigr)}{\log M} \geq -\frac{\log e}{e \log M}$$
where the right-hand side of this inequality holds since the function
$f(x) = x \log x$ for $x>0$ achieves its minimal value at $x=\frac{1}{e}$,
it follows that if the limit on the left-hand side of
\eqref{eq: condition related to the improved known bound}
is zero then also
$$\lim_{M \rightarrow \infty} \frac{\xi_{\text{avg}}^{(M)}
\log \bigl(\xi_{\text{avg}}^{(M)} \bigr)}{\log M} = 0.$$
Therefore, the definition of $K_M$ in \eqref{eq: K_M} gives that
\begin{eqnarray*}
&& \lim_{M \rightarrow \infty} \frac{\xi_{\text{avg}}^{(M)} \, \log K_M}{\log M} \\
&& = \lim_{M \rightarrow \infty} \frac{\xi_{\text{avg}}^{(M)}
\log \bigl(\xi_{\max}^{(M)} \bigr)}{\log M} - \lim_{M \rightarrow \infty} \frac{\xi_{\text{avg}}^{(M)}
\log \bigl(\xi_{\text{avg}}^{(M)} \bigr)}{\log M} \\
&& = 0.
\end{eqnarray*}
This shows that the conclusion in \eqref{eq: condition from the 1st new theorem} implies
the one in \eqref{eq: condition related to the improved known bound}.

{\em A special case of \eqref{eq: probability mass function of first example}
with numerical results}:
As a special case of the probability mass function in
\eqref{eq: probability mass function of first example},
let $M = 2^m$ for some $m \in \naturals$, let $u_i = (-1)^i$
for $i \in \{1, \ldots, M\}$, and $\xi_i = \beta$ for
some $\beta \in [0,1]$. In this special case,
\begin{eqnarray*}
P_X(a_i) = \left\{ \begin{array}{ll}
                          2^{-m} \, (1-\beta)
                          & \mbox{if $i \in \{1, 3, \ldots, 2^m-1\}$} \\[0.1cm]
                          2^{-m} \, (1+\beta)  & \mbox{if $i \in \{2, 4, \ldots, 2^m\}$.}
                          \end{array}
                          \right.
\end{eqnarray*}
Let $Y$ be a random variable that gets all the values in the set $\{a_1, \ldots, a_M\}$
with equal probability (i.e., $2^{-m}$). Then, the local and total variation distances
between $X$ and $Y$ are
$$d_{\text{loc}}(X,Y) = \frac{\beta}{M}, \quad d_{\text{TV}}(X,Y) = \frac{\beta}{2}$$
so, from \eqref{eq: alpha}, $\alpha = \frac{2}{M}$. The entropies of $X$ and $Y$ are
$$H(X) = (m-1) \log 2 + h\Bigl(\frac{1-\beta}{2}\Bigr), \quad H(Y) = m \, \log 2$$
so, $H(Y)-H(X) = \log 2 - h\bigl(\frac{1-\beta}{2}\bigr)$ independently of $m$.

For comparison, the known bound in
Theorem~\ref{theorem: known bound on the entropy difference in terms of total variation distance}
that only depends on the total variation distance between $X$ and $Y$ (with no
further knowledge about their probability mass functions) gives
$$H(Y)-H(X) \leq \frac{m \beta}{2} \cdot \log 2 + h\Bigl(\frac{\beta}{2}\Bigr) +
\frac{\beta}{2} \cdot \log(1-2^{-m})$$ so this upper bound increases almost linearly
with $m$, in contrast to the exact value that is independent of $m$. The new bound
in \eqref{eq: 1st new bound}, which depends on both the local and total variation distances
between $X$ and $Y$ (but again, without any further information on their probability mass
functions) gives
\begin{eqnarray}
&& H(Y) - H(X) \nonumber \\
&& \leq d_{\text{TV}}(X,Y) \, \log(M \alpha-1) + h\bigl(d_{\text{TV}}(X,Y)\bigr) \nonumber \\
&& = h\Bigl(\frac{\beta}{2}\Bigr).
\label{eq: bound from Theorem 4 in Example 1}
\end{eqnarray}
Similarly to the exact value, but in contrast to the former bound, the latter bound
is independent of $m$. Furthermore, if $\beta \rightarrow 0$ and $m \beta \rightarrow \infty$,
then the exact value of $H(Y)-H(X)$ as well as the latter bound (that follows from
Theorem~\ref{theorem: 1st new bound}) tend to zero, whereas the former bound that follows
from Theorem~\ref{theorem: known bound on the entropy difference in terms of total variation distance}
tends to infinity. This shows the difference in the two bounds, exemplifying
the possible advantage of taking into account the local distance in addition to the total
variation distance.

For $\beta \in [0, \frac{1}{2}]$, the condition $\frac{1}{2} \leq \frac{P_X}{P_Y} \leq 2$
is fulfilled, so the tightened bound in \eqref{eq: possible refinement of 1st new bound}
gives that
\begin{equation}
0 \leq H(Y) - H(X) \leq h\Bigl(\frac{\beta}{2}\Bigr) - \beta \, \log 2.
\label{eq: tightened bound from Theorem 4 in Example 1}
\end{equation}
If $\beta = \frac{1}{2}$, $H(Y)-H(X) = \log 2 - h\bigl(\frac{1}{2}\bigr) = 0.131$~nats,
the upper bound in \eqref{eq: bound from Theorem 4 in Example 1} is equal to
0.562~nats, and the tightened version of this bound in
\eqref{eq: tightened bound from Theorem 4 in Example 1}
is equal to 0.216~nats.

It is noted that since $P_X$ is majorized by $P_Y$ (see
\cite[Definition~1 on p.~5934]{conditional_entropy_and_error_probability_IT2010}),
then according to \cite[Theorem~3]{conditional_entropy_and_error_probability_IT2010}
$$H(Y) - H(X) \geq D(P_X || P_Y)$$
and since $P_Y$ refers to a uniform distribution over a set of cardinality $M=2^m$ then
$H(Y) = m \log 2$, and $$D(P_X || P_Y) = m \log 2 - H(P_X)$$
so, the above lower bound is achieved here with equality.
\label{example1: known probability mass function}
\end{example}

\vspace*{0.2cm}
In Example~\ref{example1: known probability mass function}, the probability
mass function of the discrete random variable $X$ was known explicitly.
However, in many interesting applications, this is not necessarily the case.
If the exact distribution of $X$ is not available or is numerically hard to compute,
a derivation of some good bounds on the local and total variation distances between
$X$ and another random variable $Y$ with a known probability mass function can be
valuable to get a rigorous bound on the difference $|H(X)-H(Y)|$ via
Theorems~\ref{theorem: 1st new bound}
or~\ref{theorem: second new theorem for the case of an infinite countable set A}.
As a result of the calculation of such a bound on the entropy difference, it
provides bounds on the entropy of $X$ in terms of another entropy (the
entropy of $Y$) which is assumed to be easily calculable. For
example, assume that $X = \sum_{i=1}^n X_i$ is expressed as a sum of Bernoulli random
variables that are either independent or weakly dependent, and may be also
non-identically distributed. Let $X_i \sim \text{Bernoulli}(p_i)$, and assume that
$\sum_{i=1}^n p_i = \lambda$ where all of the $p_i$'s are much smaller than~1.
In this case, the approximation of $X$ by a Poisson distribution with mean
$\lambda$ (according to the law of small numbers \cite{KontoyiannisHJ_2005})
raises the question: How close is $H(X)$ to the entropy of the Poisson distribution
with mean $\lambda$~? (note that the latter entropy of the Poisson distribution
is calculated efficiently in \cite{AdellLY_IT2010}). This question is especially
interesting because the support of the Poisson distribution is the infinite countable
set of non-negative integers, and the entropy is known not to be continuous when
the support is not finite; hence, a small total variation distance does not
yield in general a small difference of the two entropies.
This question was addressed in \cite{Sason_ITW} via the use of
Corollary~\ref{corollary: 2nd new corollary}, combined with an upper bound on
the total variation distance between $X$ and $Y$; the latter bound
is calculated via the use of the Chen-Stein method (see, e.g.,
\cite[Chapter~2]{RossP_book07}).

\vspace*{0.1cm}
\begin{example}[Poisson approximation]
In the following, we wish to tighten the bounds on the entropy of a sum
of independent Bernoulli random variables that are not necessarily identically
distributed. The bound provided in \cite[Corollary~1]{Sason_ITW} relies on
an upper bound on the total variation distance between this sum and a Poisson
random variable with the same mean (see \cite[Theorem~1]{BarbourH_1984} or
\cite[Theorem~2.M]{BarbourHJ_book_1992}).
In order to tighten the bound on the entropy in the considered setting,
we further rely on a new lower bound on the total variation distance
(see \cite[Theorem~1 and Corollary~1]{Sason_SPL_2013}) and an upper bound on the local
distance (see \cite[Theorem~2.Q and Corollary~9.A.2]{BarbourHJ_book_1992}). The
latter two bounds provide an upper bound on the ratio of the local and total
variation distances, which enables to apply the bound in
Theorem~\ref{theorem: second new theorem for the case of an infinite countable set A};
it improves the bound in Corollary~\ref{corollary: 2nd new corollary}
which solely relies on an upper bound on the total variation distance. It is noted
that the latter looser bound, which relies on Corollary~\ref{corollary: 2nd new corollary}
was used in \cite{Sason_ITW} for estimating the
entropy of a sum of Bernoulli random variables in the more general setting where the
summands are possibly dependent.

Let $X = \sum_{i=1}^n X_i$ be a sum of independent Bernoulli random variables such that
$X_i \sim \text{Bernoulli}(p_i)$ for every $i \in \{1, \ldots, n\}$. Let
$Y \sim \text{Po}(\lambda)$, where $\lambda = \expectation[X] = \sum_{i=1}^n p_i$, be a Poisson
random variable with mean $\lambda$. From \cite[Theorem~1]{BarbourH_1984}
(or \cite[Theorem~2.M]{BarbourHJ_book_1992}), the following upper bound on the
total variation distance holds:
\begin{equation}
d_{\text{TV}}(X, Y) \leq \left(\frac{1 - e^{-\lambda}}{\lambda}\right) \sum_{i=1}^n p_i^2.
\label{eq: upper bound on total variation distance for the Poisson approximation of independent Bernoulli summands}
\end{equation}
Furthermore, from \cite[Corollary~1]{Sason_SPL_2013}, the following lower bound on the
total variation distance holds:
\begin{equation}
d_{\text{TV}}(X,Y) \geq k \, \sum_{i=1}^n p_i^2
\label{eq: lower bound on total variation distance for the Poisson approximation of independent Bernoulli summands}
\end{equation}
where
\begin{eqnarray}
&& \hspace*{-1cm} k \triangleq \frac{e}{2 \lambda} \;
\frac{1 - \frac{1}{\theta} \, \left(3+\frac{7}{\lambda}\right)}{\theta + 2 e^{-1/2}}
\label{eq: k in the lower bound on the total variation distance} \\[0.3cm]
&& \hspace*{-1cm} \theta \triangleq 3 + \frac{7}{\lambda} + \frac{1}{\lambda} \cdot
\sqrt{(3\lambda+7)\bigl[(3+2e^{-1/2}) \lambda + 7\bigr]}.
\label{eq: theta}
\end{eqnarray}

An upper bound on the local distance between a sum of independent Bernoulli
random variables and a Poisson distribution with the same mean $\lambda$
follows as a special case of \cite[Corollary~9.A.2]{BarbourHJ_book_1992}
by setting $l=1$ (so that the distribution $Q_l$ in this corollary is specialized
for $l=1$ to the Poisson distribution $\text{Po}(\lambda)$, according to
\cite[Eq.~(1.12) on p.~177]{BarbourHJ_book_1992}). Since the upper bound on the
right-hand side of the inequality in \cite[Corollary~9.A.2]{BarbourHJ_book_1992}
does not depend on the (time) index $j$, it follows that the same bound also holds
while referring to
$$d_{\text{loc}}(X,Y) \triangleq \sup_{j \in \naturals_0} \bigl|\pr(X=j) -
\text{Po}(\lambda)\{j\}\bigr|.$$
Based on the notation used in this corollary, it implies that if
$\left(\frac{1-e^{-\lambda}}{\lambda}\right) \, \sum_{i=1}^n p_i^2 \leq \frac{1}{8}$
then the local distance between a sum of independent Bernoulli random variables
$X_i \sim \text{Bernoulli}(p_i)$ and a Poisson random variable with mean
$\lambda = \sum_{i=1}^n p_i$ is upper bounded by
\begin{eqnarray}
&& \hspace*{-1.2cm} d_{\text{loc}}(X,Y) \nonumber \\
&& \hspace*{-1.2cm} \leq 4 \, \bigl(2 \, \max_{j \in \naturals_0}
\pr(Y=j) \bigr) \left(\frac{1-e^{-\lambda}}{\lambda}\right) \, \sum_{i=1}^n p_i^2 \nonumber \\[0.1cm]
&& \hspace*{-1.2cm} \stackrel{\text(a)}{\leq} 4 \, \min \left\{\sqrt{\frac{2}{e \lambda}}, \,
2 e^{-\lambda} \, I_0(\lambda) \right\}
\left(\frac{1-e^{-\lambda}}{\lambda}\right) \, \sum_{i=1}^n p_i^2
\label{eq: simple upper bound on the local distance for the Poisson approximation
of independent summands}
\end{eqnarray}
where inequality~(a) holds due to \cite[Proposition~A.2.7 on pp.~262--263]{BarbourHJ_book_1992},
and $I_0$ denotes the modified Bessel function of order zero.
Since an upper bound on the total variation distance also forms
an upper bound on the local distance, then a combination of
\eqref{eq: upper bound on total variation distance for the Poisson approximation
of independent Bernoulli summands} and
\eqref{eq: simple upper bound on the local distance for the Poisson approximation
of independent summands} gives that
\begin{eqnarray}
d_{\text{loc}}(X,Y)
\leq \min\biggl\{\, 1, 4 \, \sqrt{\frac{2}{e \lambda}},
\, 8 e^{-\lambda} \, I_0(\lambda) \, \biggr\} \, \left(\frac{1-e^{-\lambda}}{\lambda}\right) \, \sum_{i=1}^n p_i^2.
\label{eq: tightened upper bound on the local distance for the Poisson approximation
of independent summands}
\end{eqnarray}
We now apply
Theorem~\ref{theorem: second new theorem for the case of an infinite countable set A}
to get rigorous bounds on the entropy $H(X)$ by estimating how close it is
to $H\bigl(\text{Po}(\lambda)\bigr)$. Note that the improvement in the
tightness of the bound in
Theorem~\ref{theorem: second new theorem for the case of an infinite countable set A},
in comparison to the looser bound in Corollary~\ref{corollary: 2nd new corollary},
is more significant when the ratio $\alpha$ of the local and total variation distances
is close to zero. This happens to be the case if $\lambda \gg 1$ where due to the
asymptotic expansion of $I_0$ (see \cite[Eq.~(9.7.1) on p.~377]{handbook of mathematical tables}
or \cite[Eq. (8.451.5) on p.~973]{Gradshteyn and Ryzhik})
$$I_0(\lambda) \approx \frac{e^\lambda}{\sqrt{2 \pi \lambda}} \left(1 + \frac{1}{8 \lambda}
+ \frac{9}{128 \lambda^2} + \ldots \right), \quad \mbox{if} \, \lambda \gg 1$$
one gets from Eqs.~\eqref{eq: lower bound on total variation distance for the Poisson approximation of independent Bernoulli summands}--\eqref{eq: theta} and
\eqref{eq: tightened upper bound on the local distance for the Poisson approximation
of independent summands}, combined with the limit in \cite[Eq.~(47)]{Sason_SPL_2013}, that
\begin{eqnarray}
&& \hspace*{-0.4cm} \alpha = \frac{d_{\text{loc}}(X,Y)}{d_{\text{TV}}(X,Y)} \nonumber \\
&& \hspace*{-0.4cm} \stackrel{(\text{if} \; \lambda \gg 1)}{\leq} \frac{4 \sqrt{\frac{2}{\pi \lambda}} \; \left(\frac{1-e^{-\lambda}}{\lambda}\right)
\dsum_{i=1}^n p_i^2}{\frac{e}{6} \left(1+\sqrt{1+\frac{2}{3} \cdot e^{-1/2}} \right)^{-2} \;
\left(\frac{1-e^{-\lambda}}{\lambda} \right) \dsum_{i=1}^n p_i^2} \nonumber \\
&& = \frac{24}{e} \sqrt{\frac{2}{\pi}} \left(1+\sqrt{1+\frac{2}{3} \cdot e^{-1/2}} \right)^2 \;
\sqrt{\frac{1}{\lambda}} \nonumber \\
&& \approx \frac{33.634}{\sqrt{\lambda}}
\end{eqnarray}
so, for large values of $\lambda$, the upper bound on the parameter $\alpha$ in
\eqref{eq: alpha} decays to zero like the square-root of~$\frac{1}{\lambda}$.

As a possible application, consider a noiseless binary-adder multiple-access channel
(MAC) with $n$ independent users where each user transmits binary symbols,
and the channel output is the algebraic sum of the input symbols. The
capacity region of this MAC channel is an $n$-dimensional polyhedron. One feature
of this capacity region is the sum of the rates that is given by
$R_{\text{SUM}} \triangleq \sum_{i=1}^n R_i$,
and it is upper bounded by the joint mutual information between the input symbols
$X_1, \ldots, X_n$ and the corresponding channel output $Y = \sum_{i=1}^n X_i$, i.e.,
\begin{equation*}
R_{\text{SUM}} \leq \max_{P_{\bf{X}}: P_{\bf{X}} = P_{X_1} \ldots P_{X_n}} I(X_1, \ldots, X_n; Y)
\end{equation*}
where $H(Y| X_1, \ldots, X_n) = 0$ since the MAC is noiseless and the output symbol
is the sum of the $n$ input symbols, and therefore
$I(X_1, \ldots, X_n; Y) = H(Y)$.\footnote{The reader is referred to \cite{sum_rate_noiseless_MAC_IT_1981}
for the consideration of the sum-rate for two noiseless multiple-access channels with
some similarity to the binary adder channel, see footnote in \cite[p.~43]{sum_rate_noiseless_MAC_IT_1981}.}
Hence, in the considered setting, the maximal sum rate is the maximal entropy of
the sum of $n$ independent binary random variables where $X_i \sim \text{Bernoulli}(p_i)$
for $i \in \{1, \ldots, n\}$. Under the constraint that
$\sum_{i=1}^n \expectation[X_i] \leq \lambda,$ it follows from the maximal entropy result
in \cite{Harremoes_2001}, \cite{KarlinR_81} and \cite{Shepp_Olkin_1981} that the entropy of
$Y$ is maximized when the $n$ independent inputs are i.i.d. with mean $p = \frac{\lambda}{n}$,
and consequently the channel output $Y$ is Binomially distributed with
$Y \sim \text{Binom}\bigl(n, \frac{\lambda}{n}\bigr)$. For a very large number of users,
the calculation of the entropy of the Binomial distribution is difficult, and it would be
much easier to calculate the entropy $H\bigl(\text{Po}(\lambda)\bigr)$ for a Poisson
distribution with mean $\lambda$ (see \cite{AdellLY_IT2010}).

In the following, we make use of
Theorem~\ref{theorem: second new theorem for the case of an infinite countable set A}
to get an upper bound on the entropy difference
\begin{equation}
H\bigl(\text{Po}(\lambda)\bigr) - H\Bigl(\text{Binom}\bigl(n, \frac{\lambda}{n}\bigr)\Bigr)
\label{eq: difference of the entropies of Poisson and Binomial distributions}
\end{equation}
where, due to the maximal entropy result for the Poisson distribution (see, e.g.,
\cite{Harremoes_2001}, \cite{KarlinR_81} or \cite{Shepp_Olkin_1981}), this difference
is positive. Let $X \sim \text{Binom}\bigl(n, \frac{\lambda}{n}\bigr)$ be a sum of
$n$ i.i.d. Bernoulli random variables with probability of success $p = \frac{\lambda}{n}$,
and let $Y \sim \text{Po}(\lambda)$. From
\eqref{eq: upper bound on total variation distance for the Poisson approximation of independent Bernoulli summands},
the total variation distance in this case is upper bounded by
\begin{eqnarray}
&& \hspace*{-1cm} d_{\text{TV}}(X,Y) \leq \frac{\lambda (1-e^{-\lambda})}{n} \triangleq \eta_1.
\label{eq: upper bound on the total variation distance between Poisson and Binomial distributions}
\end{eqnarray}
From \eqref{eq: lower bound on total variation distance for the Poisson approximation of
independent Bernoulli summands} and \eqref{eq: k in the lower bound on the total variation distance},
the following inequality holds:
\begin{equation}
d_{\text{TV}}(X,Y) \geq \frac{e}{2} \,
\frac{1 - \frac{1}{\theta} \, \left(3+\frac{7}{\lambda}\right)}{\theta + 2 e^{-1/2}}
\, \frac{\lambda}{n} \triangleq \eta_2
\label{eq: lower bound on the total variation distance between Poisson and Binomial distributions}
\end{equation}
where $\theta$ is given in \eqref{eq: theta}. Furthermore, for using
Theorem~\ref{theorem: second new theorem for the case of an infinite countable set A},
one needs an upper bound on the local distance between the Poisson and Binomial
distributions. Eq.~\eqref{eq: tightened upper bound on the local distance for the
Poisson approximation of independent summands} gives that
\begin{eqnarray}
d_{\text{loc}}(X,Y)
\leq \min\biggl\{\, 1, 4 \, \sqrt{\frac{2}{\pi \lambda}},
\, 8 e^{-\lambda} \, I_0(\lambda) \, \biggr\} \, \frac{\lambda \bigl(1-e^{-\lambda} \bigr)}{n} \triangleq \eta_3.
\label{eq: upper bound on the local distance between Poisson and Binomial distributions}
\end{eqnarray}
Following the notation in
Theorem~\ref{theorem: second new theorem for the case of an infinite countable set A},
it follows that $m = n+1$. From
\eqref{eq: conditions on M in the second new theorem}, one needs to choose an integer $M$
such that
\begin{equation}
M \geq \max\biggl\{n+2, \, \frac{\eta_2}{\eta_3 (1-\eta_1)} \biggr\}
\label{eq: first condition on M}
\end{equation}
and
\begin{equation}
\sum_{j=M}^{\infty} \Pi_{\lambda}(j) \leq \eta_3
\label{eq: second condition on M}
\end{equation}
where $\Pi_{\lambda}(j) \triangleq \frac{e^{-\lambda} \, \lambda^j}{j!}$ for $j \in \naturals_0$
designates the probability mass function of $\text{Po}(\lambda)$.
Based on Chernoff's bound,
\begin{eqnarray}
&& \sum_{j=M}^{\infty} \Pi_{\lambda}(j) \nonumber \\
&& = \pr(Y \geq M) \nonumber \\
&& \leq \inf_{\theta \geq 0} \left\{e^{-\theta M} \,
\expectation\bigl[e^{\theta Y}\bigr] \right\} \nonumber \\
&& = \inf_{\theta \geq 0} \left\{e^{-\theta M} \,
e^{\lambda (e^{\theta}-1)} \right\} \nonumber \\
&& = \exp\left\{ -\left[ \lambda + M \log \Bigl(\frac{M}{\lambda e}\Bigr)
\right] \right\}.
\label{eq: Chernoff inequality for Poisson distribution}
\end{eqnarray}
Let $M \geq \lambda e^2$, then it follows from \eqref{eq: second condition on M}
and \eqref{eq: Chernoff inequality for Poisson distribution} that it is sufficient
for $M$ to satisfy the condition $$\exp\bigl(-(\lambda+M)\bigr) \leq \eta_3.$$ Combining
it with \eqref{eq: first condition on M} leads to the following possible choice of $M$:
\begin{equation}
M \triangleq \max \biggl\{n+2, \, \frac{\eta_2}{\eta_3 (1-\eta_1)}, \,
\lambda e^2, \, \log \biggl(\frac{1}{\eta_3}\biggr) - \lambda \biggr\}
\label{eq: M for the binary-adder MAC}
\end{equation}
where $\eta_1$, $\eta_2$ and $\eta_3$ are introduced in
\eqref{eq: upper bound on the total variation distance between Poisson and Binomial distributions},
\eqref{eq: lower bound on the total variation distance between Poisson and Binomial distributions},
and \eqref{eq: upper bound on the local distance between Poisson and Binomial distributions}
respectively. Finally, for the use of
Theorem~\ref{theorem: second new theorem for the case of an infinite countable set A}, one
needs to choose $\eta_4 > 0$ such that
$\sum_{j=M}^{\infty} \bigl\{-\Pi_{\lambda}(j) \; \log \bigl(\Pi_{\lambda}(j)\bigr) \bigr\}
\leq \eta_4.$ Straightforward calculation gives that
\begin{eqnarray}
&& \sum_{j=M}^{\infty} \bigl\{-\Pi_{\lambda}(j) \, \log \Pi_{\lambda}(j) \bigr\} \nonumber \\
&& = -\lambda \, \log \lambda \sum_{j=M-1}^{\infty} \Pi_{\lambda}(j)
+ \lambda \sum_{j=M}^{\infty} \Pi_{\lambda}(j)
+ \sum_{j=M}^{\infty} \Pi_{\lambda}(j) \, \log(j!) \, .
\label{eq: bound on the second sum for the derivation of the error bound on the entropy - 1st step}
\end{eqnarray}
From Stirling's formula, the
equality $j ! = \sqrt{2\pi j} \left(\frac{j}{e}\right)^j \, e^{\xi_j}$
holds for every $j \in \naturals$ and for some
$\xi_j \in \bigl(\frac{1}{12j+1}, \frac{1}{12j}\bigr)$. This therefore implies that
the third infinite sum on the right-hand side of
\eqref{eq: bound on the second sum for the derivation of the error bound on the entropy - 1st step}
satisfies
\begin{eqnarray}
&& \sum_{j=M}^{\infty} \Pi_{\lambda}(j) \, \log(j!) \nonumber \\
&& \leq \sum_{j=M}^{\infty} \Pi_{\lambda}(j) \, \log\left(\sqrt{2\pi j}
\left(\frac{j}{e}\right)^j \, e^{\frac{1}{12j}} \right) \nonumber \\
&& = \frac{\log(2\pi)}{2} \sum_{j=M}^{\infty} \Pi_{\lambda}(j) +
\sum_{j=M}^{\infty} \Pi_{\lambda}(j) \left[\bigl(j+\frac{1}{2}\bigr) \log(j)-j \right]
+ \frac{1}{12} \sum_{j=M}^{\infty} \frac{\Pi_{\lambda}(j)}{j} \nonumber \\
&& \leq \frac{\log(2\pi)}{2} \sum_{j=M}^{\infty} \Pi_{\lambda}(j) +
\sum_{j=M}^{\infty} \bigl\{j(j-1) \, \Pi_{\lambda}(j) \bigr\}
+ \frac{1}{12} \sum_{j=M}^{\infty} \Pi_{\lambda}(j) \nonumber \\
&& \stackrel{\text{(a)}}{=} \frac{\log(2\pi)}{2} \sum_{j=M}^{\infty} \Pi_{\lambda}(j) +
\lambda^2 \sum_{j=M-2}^{\infty} \Pi_{\lambda}(j)
+ \frac{1}{12} \sum_{j=M}^{\infty} \Pi_{\lambda}(j) \nonumber \\
&& \leq \left( \frac{6\log(2\pi)+1}{12} + \lambda^2 \right) \sum_{j=M-2}^{\infty} \Pi_{\lambda}(j)
\label{eq: bound on the second sum for the derivation of the error bound on the entropy - 2nd step}
\end{eqnarray}
where equality~(a) follows from the identity
$$j(j-1) \, \Pi_{\lambda}(j) = \lambda^2 \, \Pi_{\lambda}(j-2), \quad \forall \, j \geq 2.$$
By combining \eqref{eq: bound on the second sum for the derivation of the error bound
on the entropy - 1st step} and
\eqref{eq: bound on the second sum for the derivation of the error bound on the
entropy - 2nd step}, it follows that
\begin{eqnarray*}
&& \sum_{j=M}^{\infty} \bigl\{ -\Pi_{\lambda}(j) \, \log \Pi_{\lambda}(j) \bigr\}
\nonumber \\
&& \leq \left(\lambda \, \log\Bigl(\frac{e}{\lambda}\Bigr)\right)_{+} \,
\sum_{j=M-1}^{\infty} \Pi_{\lambda}(j)
+ \left( \frac{6\log(2\pi)+1}{12} + \lambda^2 \right) \sum_{j=M-2}^{\infty} \Pi_{\lambda}(j) \nonumber \\
&& \leq \left[\left(\lambda \, \log\Bigl(\frac{e}{\lambda}\Bigr)\right)_{+} +
\lambda^2 + \frac{6\log(2\pi)+1}{12} \right] \sum_{j=M-2}^{\infty} \Pi_{\lambda}(j)
\end{eqnarray*}
where $M$ is introduced in \eqref{eq: M for the binary-adder MAC}, and $(x)_+ \triangleq
\max\{x, 0\}$ for every $x \in \reals$. From \eqref{eq: Chernoff inequality for Poisson distribution}
and the last inequality, it follows that $\eta_4$ can be chosen to be
\begin{eqnarray}
&& \eta_4 \triangleq \left[ \Bigl(\lambda
\log \Bigl(\frac{e}{\lambda}\Bigr)\Bigr)_+ \,
+ \lambda^2 + \frac{6 \log(2\pi) + 1}{12} \right]
\cdot \exp \left\{-\left[\lambda + (M-2) \, \log\left(\frac{M-2}{\lambda e} \right)
\right] \right\}.
\label{eq: upper bound on the tail of the entropy for the Poisson distribution}
\end{eqnarray}
At this stage, we are ready to apply
Theorem~\ref{theorem: second new theorem for the case of an infinite countable set A}
to derive a bound on the non-negative difference of the entropies in
\eqref{eq: difference of the entropies of Poisson and Binomial distributions}.
From Theorem~\ref{theorem: second new theorem for the case of an infinite countable set A},
it follows that
\begin{eqnarray}
&& \hspace*{-0.3cm} 0 \leq H\bigl(\text{Po}(\lambda)\bigr) - H\Bigl(\text{Binom}\bigl(n, \frac{\lambda}{n}\bigr)\Bigr) \nonumber \\
&& \leq \eta_1 \, \log \biggl(\frac{M \eta_3}{\eta_2} - 1 \biggr) + h(\eta_1) + \eta_4.
\label{eq: upper bound on the difference of the entropies of Poisson and Binomial
distributions}
\end{eqnarray}
For comparison, it follows from Corollary~\ref{corollary: 2nd new corollary} that the
upper bound on the right-hand side of
\eqref{eq: upper bound on the difference of the entropies of Poisson and Binomial
distributions} is replaced by
\begin{equation}
\eta_1 \, \log(\tilde{M}-1) + h(\eta_1) + \eta_4
\label{eq: weaker upper bound on the difference of the entropies of Poisson and Binomial
distributions}
\end{equation}
where
\begin{equation}
\tilde{M} \triangleq \max\Bigl\{n+2, \, \frac{1}{1-\eta_1}\Bigr\}.
\label{eq: modified M in the case that the second corollary is used}
\end{equation}
Note that the bound in
\eqref{eq: upper bound on the difference of the entropies of Poisson and Binomial
distributions} improves the bound in
\eqref{eq: weaker upper bound on the difference of the entropies of Poisson and Binomial distributions}
if $\eta_3 < \eta_2$ (i.e., if the upper bound on the
local distance is smaller than the lower bound on the total variation distance). Furthermore, the
latter bound does not take into account the parameters $\eta_2$ and $\eta_3$. As a numerical
example, for $n=10^6$ and $p=0.1$, lets check the bound on the entropy difference in
\eqref{eq: difference of the entropies of Poisson and Binomial distributions} for
$\lambda = np$ (i.e., $\lambda=10^5$).
Eqs.~\eqref{eq: upper bound on the total variation distance between Poisson and Binomial distributions}--\eqref{eq: upper bound on the local distance between Poisson and Binomial distributions},
\eqref{eq: M for the binary-adder MAC},
\eqref{eq: upper bound on the tail of the entropy for the Poisson distribution} and
\eqref{eq: modified M in the case that the second corollary is used} yield that
\begin{eqnarray*}
&& \eta_1 = 10^{-1}, \; \eta_2 = 9.5 \cdot 10^{-3}, \; \eta_3 = 1.0 \cdot 10^{-3},
\; \eta_4 \approx 0, \\
&& M=\tilde{M}=10^6+2
\end{eqnarray*}
and the two bounds in
\eqref{eq: upper bound on the difference of the entropies of Poisson and Binomial
distributions} and
\eqref{eq: weaker upper bound on the difference of the entropies of Poisson and Binomial
distributions} are, respectively, equal to 1.483 and 1.707 nats, respectively.
The value of $H\bigl(\text{Po}(\lambda)\bigr)$ is 7.175 nats, so the entropy
$H\bigl(\text{Binom}(n, \frac{\lambda}{n})\bigr)$ ranges between 5.693 to 7.175~nats. Note
that for $n=10^6$ and $\lambda=10^4$, where $p = \frac{\lambda}{n}$ is decreased
from $10^{-1}$ to $10^{-2}$, the upper bounds on
\eqref{eq: difference of the entropies of Poisson and Binomial distributions} are
decreased, respectively, to 0.183 and 0.194~nats, and $H\bigl(\text{Po}(\lambda)\bigr)
= 6.024$~nats. The Poisson approximation is more accurate in the latter case,
consistently with the law of small numbers (see, e.g., \cite{KontoyiannisHJ_2005}).
\label{example2: Poisson approximation for independent summands}
\end{example}

\begin{remark}
Example~\ref{example2: Poisson approximation for independent summands}
considers the use of
Theorem~\ref{theorem: second new theorem for the case of an infinite countable set A}
for the estimation of the entropy of a sum of independent Bernoulli random
variables. The more general case of the estimation of the entropy (via rigorous bounds)
for a sum of possibly dependent Bernoulli random variables was considered in
\cite{Sason_ITW} by using the
looser bound in Corollary~\ref{corollary: 2nd new corollary} with an upper
bound on the total variation distance that follows from the Chen-Stein method (see
\cite[Theorem~1]{ArratiaGG_AOP}). It is noted that, in principle, also the sharper bound
in Theorem~\ref{theorem: second new theorem for the case of an infinite countable set A}
can be applied to obtain bounds on the entropy for a sum of possibly dependent Bernoulli
random variables. To this end, in addition to the upper bound on the total variation
distance in \cite[Theorem~1]{ArratiaGG_AOP}, one needs to rely on a lower bound
on the total variation distance (see \cite[Chapter~3]{BarbourHJ_book_1992}) and an
upper bound on the local distance (see \cite[Theorem~2.Q on p.~42]{BarbourHJ_book_1992}). It
is noted, however, that these distance bounds are much simplified in the setting of
independent summands (see Example~\ref{example2: Poisson approximation for independent summands}).
\label{remark: dependent summands of Bernoulli random variables - the Poisson approximation}
\end{remark}

\begin{remark} The Chen-Stein method for the Poisson approximation was adapted in
\cite{Geometric approximation - 1996} to the setting of the geometric
distribution, and it yields a convenient method for assessing the accuracy
of the geometric approximation to the distribution of the number of failures
preceding the first success in dependent trials. A recent study of upper bounds
on the total variation and local distances for the geometric approximation
(respectively, denoted by $d_1$ and $d_2$ in \cite{Geometric_approximation - 2012})
enables to apply the entropy bounds in
Theorem~\ref{theorem: second new theorem for the case of an infinite countable set A}
and Corollary~\ref{corollary: 2nd new corollary} in a conceptually similar way to
Example~\ref{example2: Poisson approximation for independent summands}.
Furthermore, the entropy bound in
Corollary~\ref{corollary: 2nd new corollary} can be applied to compound geometric
and negative binomial approximations, based on upper bounds on the total variation
distance that were derived via Stein's method in \cite{compound geometric approximation}
and \cite{negative binomial approximation}, respectively.
\label{remark: geometric approximation}
\end{remark}

\section{Summary and Outlook}
\label{section: summary}
This paper is motivated by the fundamental question of quantifying the continuity
(or lack of it) of entropy, with respect to natural topologies on discrete
probability distributions. This question has been studied in the literature for the
topology induced by the total variation distance, and there it is well known
that the entropy is continuous when the alphabet is finite, but
not when the alphabet is countably infinite (see,
e.g., \cite{entropy_difference_and_variational_distance_IT2010} and
references therein). To set terminology, the local and total variation
distances are introduced in Definition~\ref{definition: local and total variation distances}
(see Section~\ref{section: Introduction}): the local distance between
two discrete random variables is defined to be the $l^{\infty}$ distance
between their probability mass functions, and the total variation distance
is half the $l^1$ distance; it is easy to show that the local distance is
less than or equal to the total variation distance.

A key tool in this paper is an explicit construction for maximal coupling,
i.e., a coupling $(\hat{X}, \hat{Y})$ of the random variables $X$ and $Y$
that maximizes the probability $\pr(\hat{X} = \hat{Y})$.
The notion of {\em maximal coupling} is also known to
be useful for the derivation of error bounds via Stein's method (see, e.g.,
\cite[Chapter~2]{RossP_book07} and \cite{Ross_Tutorial11}). Stein's method also serves
in this paper to exemplify the use of the new bounds in the context of the Poisson approximation
\cite{BarbourHJ_book_1992};
this is done by using good upper and lower bounds on the total variation distance (see
\cite{BarbourH_1984} and \cite{Sason_SPL_2013}) and a good upper bound on the
local distance \cite{BarbourHJ_book_1992}.
The link between Stein's method and information theory was pioneered in \cite{BarbourJKM_EJP_2010}
in the context of the compound Poisson approximation, and it has been further recently studied
in \cite{Ley_Swan_arxiv2012}.

This paper starts by introducing preliminary material in Sections~\ref{section: Introduction}
and~\ref{section: A Proof of a Known Bound on the Entropy of Discrete Random Variables via Coupling};
Theorems~\ref{theorem: maximal coupling}--\ref{theorem: known bound on the entropy difference in terms of total variation distance} are known results on maximal coupling, and a bound from \cite{Zhang_IT2007} on the
difference of the entropies of two discrete random variables in terms of the total variation distance.
Note that the proofs of these known results are important for the analysis in this paper.

The new results in this paper are the following:
\begin{enumerate}
\item For two given distributions on a finite alphabet, if the local distance
is strictly smaller than the total variation distance, then Theorem~\ref{theorem: 1st new bound}
provides a new bound which can be significantly better than the previously best known
bound (Theorem~\ref{theorem: known bound on the entropy difference in terms of total variation distance})
due to Zhang \cite{Zhang_IT2007}.
\item For countably infinite alphabets, a knowledge of the total variation distance
between two distributions is not sufficient for establishing an informative bound
on the difference of entropies (i.e., one has discontinuity of entropy).
Theorem~\ref{theorem: second new theorem for the case of an infinite countable set A}
demonstrates that if one of the distributions is finitely supported and some knowledge of
the other distribution is available, then the knowledge of the local and total variation
distances (or bounds on these distances) allows one to bound the difference of the entropies
even in this case.
\item Refined bounds on the entropy of near-uniform random variables on large alphabets,
as well as of sums of independent Bernoulli random variables (which arise in numerous
applications, see \cite{ArratiaGG_AOP} and references therein) are obtained in Section~\ref{section: Examples}
(see Examples~\ref{example1: known probability mass function}
and~\ref{example2: Poisson approximation for independent summands}). These refined bounds
are compared with previously known bounds. One special case
where the entropy can be explicitly evaluated and compared to various bounds is worked out,
and it is shown that Theorem~\ref{theorem: 1st new bound} improves significantly the known
bound in Theorem~\ref{theorem: known bound on the entropy difference in terms of total variation distance}.
\end{enumerate}
\vspace*{0.2cm}
A natural question that arises in the context of this paper is what if one only has bounds
on the local distance~? A treatment of this problem (which does not exist in the literature)
possibly gives further insight into why the local distance is useful in combination of the
total variation distance. In the finite alphabet case, the two metrics are equivalent since
$$d_{\text{loc}}(X,Y) \leq d_{\text{TV}}(X,Y) \leq \frac{M}{2} \cdot d_{\text{loc}}(X,Y)$$
and, hence, generate the same topology; so the bare continuity of entropy is guaranteed for
finite alphabets, and so is the discontinuity of the entropy for infinite alphabets. But are
there tight bounds on the difference of entropies just based on the local distance for finite
alphabets ?

The following proposition suggests a simple bound on the difference of entropies of two
discrete random variables that are finitely supported, based only on their local distance:
\begin{proposition}
Let $X$ and $Y$ be discrete random variables that take values in a finite set $\mathcal{A}$,
and let $|\mathcal{A}| = M$. If $d_{\text{loc}}(X,Y) \leq \frac{1}{e}$, then
\begin{eqnarray}
|H(X) - H(Y)| \leq -M \, d_{\text{loc}}(X,Y) \, \log\bigl( d_{\text{loc}}(X,Y) \bigr)
\label{eq: bound on difference of entropies based on local distance for finite alphabet}
\end{eqnarray}
with the convention that $0 \log 0$ means~0.
\label{proposition: bound on difference of entropies based on local distance for finite alphabet}
\end{proposition}
\begin{proof}
The derivation of this bound forms a small modification of the proof of \cite[Theorem~17.3.3]{Cover_Thomas}.
Let $P_X$ and $P_Y$ denote the probability mass functions of $X$ and $Y$, respectively, and let
$r(u) \triangleq | P_X(u) - P_Y(u) |$ for every $u \in \mathcal{A}$. From
\cite[Eqs.~(17.27)--(17.30)]{Cover_Thomas}, if $r(u) \leq \frac{1}{2}$ for every $u \in \mathcal{A}$,
then
$$ |H(X)-H(Y)| \leq \sum_{u \in \mathcal{A}} -r(u) \, \log r(u).$$
The bound in \eqref{eq: bound on difference of entropies based on local distance for finite alphabet}
now follows from the simple inequality $r(u) \leq d_{\text{loc}}(X,Y)$ for every $u \in \mathcal{A}$
(by definition), and due to the fact that the function $f(x) = -x \log(x)$ is monotonic increasing
over the interval $[0, \frac{1}{e}]$.
\end{proof}

\begin{remark}
The bound in \eqref{eq: bound on difference of entropies based on local distance for finite alphabet}
does not necessarily hold if $d_{\text{loc}}(X,Y) > \frac{1}{e}$. As a counter example, let $\mathcal{A}$
be a set of 3 elements, and let
$$P_X = \Bigl(\frac{1}{2}, \frac{1}{2}, 0\Bigr), \quad P_Y = (0,0,1).$$
Then $d_{\text{loc}}(X,Y)=1$ and $H(X)-H(Y)=\log 2$, so
\eqref{eq: bound on difference of entropies based on local distance for finite alphabet}
is not satisfied due to the violation of the condition on the local distance in
Proposition~\ref{proposition: bound on difference of entropies based on local distance for finite alphabet}.
\end{remark}

\vspace*{0.1cm}
\begin{remark}
A slight loosening of the bound in
\eqref{eq: bound on difference of entropies based on local distance for finite alphabet}
gives that if $d_{\text{loc}}(X,Y) \leq \frac{1}{e}$, then
$$|H(X) - H(Y)| \leq M \, h\bigl(d_{\text{loc}}(X,Y)\bigr)$$
where $h$ is the binary entropy function. In the simple case where the probability mass functions
of $X$ and $Y$ are equal to $P_X = (1-\varepsilon, \varepsilon)$ and $P_Y = (1,0)$, respectively,
we have $d_{\text{loc}}(X,Y) = \varepsilon$;
if $0 < \varepsilon \leq \frac{1}{e}$, the bound on $|H(X)-H(Y)|$ is twice larger than its exact
value that is equal to $h(\varepsilon).$ Even in this simple case, the bound on the difference
of the entropies that only depends on the local distance is not tight. On one hand, it will be of interest
to derive tighter bounds on the difference of entropies for finite
alphabets that are just based on the local distance; on the other hand, even the simple
bound in Proposition~\ref{proposition: bound on difference of entropies based on local distance for finite alphabet}
provides some insight into why the local distance is useful in combination of the
total variation distance for upper bounding the difference of entropies for finite
alphabets (see Theorem~\ref{theorem: 1st new bound}).
\end{remark}

\subsection*{Acknowledgment}
An anonymous reviewer of this journal paper and the conference version at ISIT~2013
is gratefully acknowledged for suggestions that led to an improvement of the
presentation, and for raising the question in Section~\ref{section: summary}
that led to the bound in
Proposition~\ref{proposition: bound on difference of entropies based on local distance for finite alphabet}.
The Associate Editor, Ioannis Kontoyiannis, is acknowledged for handling the manuscript.
This research work was supported by the Israeli Science Foundation (ISF), grant number 12/12.

\end{document}